\documentclass[preprint,showpacs,preprintnumbers,amsmath,amssymb]{revtex4}

\usepackage{graphicx}% Include figure files
\usepackage{dcolumn}% Align table columns on decimal point
\usepackage{bm}% bold math
\usepackage{float,color}

\begin{document}

\title{``Sloppy" nuclear energy density functionals: effective model reduction}

\author{Tamara Nik\v{s}i\'{c} and Dario Vretenar}
\affiliation{Physics Department, Faculty of Science, University of Zagreb, 10000 Zagreb, Croatia}

\date{\today}

\begin{abstract} 
Concepts from information geometry are used to analyse parameter sensitivity for a nuclear energy density 
functional, representative of a class of semi-empirical functionals that start from a microscopically motivated 
ansatz for the density dependence of the energy of a system of protons and neutrons. It is shown that such functionals 
are ``sloppy", namely characterised by an exponential range of sensitivity to parameter variations. Responsive 
to only a few stiff parameter combinations, sloppy functionals exhibit an exponential decrease of sensitivity to variations of 
the remaining soft parameters. By interpreting the space of model predictions as a manifold embedded in the data 
space, with the parameters of the functional as coordinates on the manifold, it is also shown that the exponential distribution 
of model manifold widths corresponds to the range of parameter sensitivity. Using the Manifold Boundary 
Approximation Method, we illustrate how to systematically construct effective nuclear density functionals of successively 
lower dimension in parameter space until sloppiness is eventually eliminated and the resulting functional contains only 
stiff combinations of parameters.
\end{abstract}

\pacs{21.60.Jz, 02.40.Sf, 05.10.-a}

\maketitle

%============================================================
%  Section 1
\section{\label{sec-introduction}Introduction}
%============================================================

Over the last two decades nuclear energy density functionals (NEDF), and structure models based on them, 
have become the tool of choice for the description of ground-state properties and low-energy collective excitation 
spectra of medium-heavy and heavy nuclei. A variety of structure phenomena have been successfully analysed 
using the NEDF framework, from clustering in relatively light nuclei to the stability of superheavy systems, and 
from bulk and spectroscopic properties of stable nuclei to the physics of exotic nuclei at the particle drip lines. 
In nuclear structure physics no other theoretical method can achieve the same level of global precision and 
accuracy over the entire chart of nuclides, at a comparable computational cost. 

The unknown exact nuclear EDF is approximated by functionals of powers and gradients of 
ground-state nucleon densities and currents, representing distributions of matter, spin, isospin, momentum 
and kinetic energy. A generic density functional is not necessarily related to the underlying 
inter-nucleon interactions and, in fact, some of the most successful functionals are entirely empirical.
A major long-term goal, however, is to build a fully microscopic foundation for a universal EDF framework, 
that is, an {\em ab initio} approach based directly on 
a microscopic nuclear Hamiltonian that describes two-nucleon and few-body scattering and 
bound-state observables. Important advances have been reported in the derivation of 
microscopic constraints on the analytical form of the functional and the values of its couplings 
from many-body perturbation theory starting from the underlying two- and three-nucleon interactions, 
as well as in establishing the connection between microscopic EDF methods with {\em ab initio} 
many-body techniques applicable to light nuclei \cite{DFP.10,KW.10,Geb.10,Sto.10,HKW.11,Bog.11,GCP.11,DBR.12,HKW.13,Dob.16}.

An alternative and relatively simpler approach considers semi-empirical functionals that start form a microscopically motivated 
ansatz for the nucleonic density dependence of the energy of a system of protons and neutrons. Most of the parameters 
of such a functional are adjusted, in a local density approximation, to reproduce a given microscopic equation of state of 
infinite symmetric and asymmetric nuclear matter, and eventually neutron matter. Parameters that correspond to derivative terms 
can be determined, in a generalised gradient approximation, from microscopic calculations of inhomogeneous or semi-infinite 
nuclear matter. The remaining, usually few, terms that do not contribute to the energy density at the nuclear matter level, are then 
adjusted to selected ground-state data of an arbitrarily large set of spherical and/or deformed nuclei. A number of semi-empirical 
functionals have been developed over the last decade
\cite{TW.99,Fin.04,Fin.06,Baldo.08,Baldo.13,NVR.08,NVR.11,Kor.10,Kor.12,Kor.14,Bul.15}, and very successfully applied to 
studies of a diversity of structure properties of the entire chart of nuclides. 

There are, of course, also dozens of fully phenomenological functionals and effective mean-field interactions that have been 
adjusted and analysed over almost forty years. Most of these functionals make use of some properties of nuclear 
matter at saturation, such as the saturation density, binding energy at saturation, compression modulus, symmetry energy, 
etc., but this input is not based on microscopic many-body calculations, rather it is empirical. All the parameters, with a possible 
exception of few determined by an educated guess, are usually adjusted in a least-square fit to empirical properties of nuclear matter 
at saturation and ground-state data of finite nuclei. A variety of Skyrme, Gogny, and relativistic 
EDFs have been employed, particularly in the 
last decade \cite{BHR.03,SR.07,EKR.11,GCP.09,GCP.13,BGG.89,BGG.91,CGH.08,GHGP.09,CPGB.15,EDF.04,VALR.05,Meng.16}, 
to explore many nuclear properties, from masses, radii and deformations, to modes of collective excitation and rotational bands, and 
also been used in astrophysical applications. Based on these functionals, structure models have been developed that extend 
the framework beyond the self-consistent mean-field, and are currently employed in spectroscopic studies of excitation spectra and 
decay rates. 
 
One of most serious problems with the development of the NEDF framework is that it is difficult to compare results obtained 
with different functionals, either because they significantly differ in the functional dependence on the nucleonic density, or
because they include different subsets of terms of a general functional characteristic for a certain class, that is, microscopic, 
semi-empirical or phenomenological. Ideally, model dependence could be removed by including all terms allowed by 
symmetries. However, available data can only constrain a relatively small subset of terms in a general expansion, and 
additional criteria are required for selecting the optimal energy density functional form.

Recently a series of studies have been initiated on uncertainty quantification and propagation of errors in nuclear 
Density Functional Theory. Using statistical methods and advanced computing techniques, the stability and 
interdependence of parameters that determine various functionals, the inherent parameter uncertainties and 
their propagation, and the corresponding uncertainties of 
predicted observables have been analysed 
\cite{Dob14a,Rei10,Fat11,Gao.13,Chen.14,GC.14,PCF.15,ER.15,Nik.15,Sch.15,Sch.15a,Rei.16,Uta.16}. 
Many new and useful results have been obtained, and some common characteristics have become apparent. 
Nuclear energy density functionals, with typically ten or more adjustable parameters depending on the specific 
application, display great flexibility but essentially depend on just a few stiff parameter combinations that 
can be tightly constrained by the underlying microscopic theory and/or experimental observations. Predicted 
results generally display accurate interpolations in known regions of the nuclide chart. However, predictions from 
various models tend to diverge when extrapolated to areas where data are insufficient. More importantly, 
NEDFs often exhibit an exponential range of sensitivity to parameter variations, and one finds many
soft linear combinations of bare model parameters that are poorly constrained by data. This characteristic, however, 
does not necessarily correspond to a trivial case of too many parameters adjusted to insufficient data. It might actually point to  
the presence of low-dimensional effective functionals associated with the few stiff parameter combinations. 

This general property shared by many effective macroscopic models of complex but unresolved 
microscopic degrees of freedom, also referred to as ``sloppines'', occurs in 
many fields of science. The interesting problem of a systematic construction of reduced 
low-dimensional models from a complete but sloppy framework of much higher dimension has recently 
attracted considerable interest in diverse areas of physics, chemistry, biology, medicine, etc. (see, for instance, 
Refs.~\cite{Machta.13,Transtrum.15,Buch.15}). In this work concepts from information 
geometry are applied to analyse sloppiness of semi-empirical nuclear energy density functionals and, 
in particular, a recently introduced method \cite{Transtrum.14} is employed to reduce a sloppy 
functional to a simpler effective functional of lower dimension in parameter space. As an illustrative example we 
consider the relativistic density functional DD-PC1 \cite{NVR.08}, which has successfully been used in studies 
of many nuclear properties, both at the self-consistent mean-field level, as well as a basis for recent  
beyond-mean-field spectroscopic calculations. The functional is introduced and briefly discussed in 
Sec.~\ref{sec-functional}. Section \ref{sec-theoretical} contains the theoretical framework of the present 
analysis, and describes the fit of the parameters of the functional to a microscopic equation of state of 
symmetric nuclear matter. The reduction of the parameter space dimension and the resulting 
simplification of the functional density dependence by the Manifold Boundary Approximation 
Method \cite{Transtrum.14} is described in Sec.~\ref{MBAM}. A brief summary of the main results and 
an outlook for future studies are included in Sec.~\ref{sec-summary}.
\newpage
%============================================================
%  Section 2
\section{\label{sec-functional}  The Relativistic Density Functional DD-PC1}
%============================================================
A relativistic nuclear energy density functional that explicitly includes nucleon degrees of freedom only, 
can be expressed in terms of  the densities and currents
bilinear in the Dirac spinor field $\psi$ of the nucleon:
$\bar{\psi}\mathcal{O}_\tau \Gamma \psi$, $\mathcal{O}_\tau \in \{1,\tau_i\}$, 
   $\Gamma \in \{1,\gamma_\mu,\gamma_5,\gamma_5\gamma_\mu,\sigma_{\mu\nu}\}$.
Here $\tau_i$ are the isospin Pauli matrices and $\Gamma$ generically denotes
the Dirac matrices. The ground-state density and energy of a nucleus are then determined by the
self-consistent solution of the corresponding relativistic single-nucleon Kohn-Sham equations. 
In lowest order the interaction Lagrangian contains four-fermion
(contact) terms in the various isospace-space channels:
\begin{center}
\begin{tabular}{ll}
 isoscalar-scalar:   &   $(\bar\psi\psi)^2$\\
 isoscalar-vector:   &   $(\bar\psi\gamma_\mu\psi)(\bar\psi\gamma^\mu\psi)$\\
 isovector-scalar:  &  $(\bar\psi\vec\tau\psi)\cdot(\bar\psi\vec\tau\psi)$\\
 isovector-vector:   &   $(\bar\psi\vec\tau\gamma_\mu\psi)
                         \cdot(\bar\psi\vec\tau\gamma^\mu\psi)$ .\\
\end{tabular}
\end{center}
Vectors in isospin space are denoted by arrows. A more general Lagrangian
can be written as a power series in the currents
 $\bar{\psi}\mathcal{O}_\tau\Gamma\psi$ and their derivatives, with higher-order
terms representing in-medium many-body correlations or, in an alternative approach 
that directly leads to  linear single-nucleon Kohn-Sham equations,
the Lagrangian is constructed with second-order
interaction terms only, with many-body correlations encoded in
density-dependent coupling functions. In complete analogy to the
successful meson-exchange relativistic mean-field phenomenology, in which
the isoscalar-scalar sigma-meson, the isoscalar-vector omega-meson,
and the isovector-vector rho-meson build the minimal set of meson fields that is
necessary for a quantitative description of nuclei,
an effective Lagrangian that includes the isoscalar-scalar, isoscalar-vector and
isovector-vector four-fermion interactions reads:
\begin{align}
\label{Lagrangian}
\mathcal{L} &= \bar{\psi} (i\gamma \cdot \partial -M)\psi \nonumber \\
     &- \frac{1}{2}\alpha_s({\rho})(\bar{\psi}\psi)(\bar{\psi}\psi)
       - \frac{1}{2}\alpha_v({\rho})(\bar{\psi}\gamma^\mu\psi)(\bar{\psi}\gamma_\mu\psi)
     - \frac{1}{2}\alpha_{tv}({\rho})(\bar{\psi}\vec{\tau}\gamma^\mu\psi)
                                                                 (\bar{\psi}\vec{\tau}\gamma_\mu\psi) \nonumber \\
    &-\frac{1}{2} \delta_S (\partial_\nu \bar{\psi}\psi)  (\partial^\nu \bar{\psi}\psi)
         -e\bar{\psi}\gamma \cdot A \frac{(1-\tau_3)}{2}\psi\;.
\end{align}
In addition to the free-nucleon Lagrangian and the point-coupling interaction terms,
when applied to nuclei the model must include the coupling of the protons to the
electromagnetic field. The couplings $\alpha_s$, $\alpha_v$, and $\alpha_{tv}$ 
are functionals of the nucleon one-body density $\rho$. 
The derivative term in Eq.~(\ref{Lagrangian}), with a single constant parameter, 
accounts for leading effects of finite-range interactions that are crucial for a 
quantitative description of nuclear density distributions, e.g. nuclear radii. This Lagrangian does not
include isovector-scalar terms, that is, the channel that in the meson-exchange
picture is represented by the exchange of an effective $\delta$-meson. The reason is that, 
although the functional in the isovector channel can be constrained by the 
nuclear matter symmetry energy and data on finite nuclei, the partition between the 
scalar and vector isovector terms is not determined by ground-state data. 

To specify the medium dependence of the couplings
$\alpha_s$, $\alpha_v$, and $\alpha_{tv}$, one
could start from a microscopic (relativistic) equation of state (EoS) of symmetric
and asymmetric nuclear matter, and map the corresponding nucleon self-energies
onto the mean-field self-energies that determine the single-nucleon Dirac equation 
(local density approximation).
However, energy density functionals fully determined directly by a microscopic EoS do not 
reproduce data in finite nuclei with sufficient accuracy. A fully 
phenomenological approach, on the other hand,  
starts from an assumed ansatz for the medium dependence of the mean-field
nucleon self-energies, and adjusts the model parameters directly to nuclear data.
In a semi-empirical approximation, 
guided by the microscopic density dependence of the vector and scalar
self-energies, the following practical ansatz for the functional form
of the coupling parameters in Eq.~(\ref{Lagrangian}) was adopted in Ref.~\cite{NVR.08}:
\begin{eqnarray}
\alpha_s(\rho)&=& a_s + (b_s + c_s x)e^{-d_s x},\nonumber\\
\alpha_v(\rho)&=& a_v +  b_v e^{-d_v x},\label{parameters} \\
\alpha_{tv}(\rho)&=& b_{tv} e^{-d_{tv} x},\nonumber
\end{eqnarray}
with $x=\rho/\rho_\mathrm{sat}$, where $\rho_\mathrm{sat}$ indicates the nucleon
density at saturation in symmetric nuclear matter.  The set of 10
parameters was adjusted in a {\em least-squares} fit
to the binding energies of 64 axially deformed nuclei
in the mass regions $A\approx 150-180$ and $A\approx 230-250$. The
resulting functional DD-PC1 \cite{NVR.08} has been further tested in
calculations of binding energies, charge radii, deformation
parameters, neutron skin thickness, and excitation energies of giant
monopole and dipole resonances. It has also been successfully applied in a number of 
beyond-mean-field studies of spectroscopic properties based on the generator coordinate 
method and the quadrupole (octupole) 
collective Hamiltonian \cite{NVR.11,LNV.16}. The nuclear matter equation of state that 
corresponds to DD-PC1 is characterised by the following values of pseudo-observables
at the saturation point: nucleon density
$\rho_\mathrm{sat}=0.152~\textnormal{fm}^{-3}$, volume energy $a_v=-16.06$
MeV, surface energy $a_s=17.498$ MeV, symmetry energy $a_4 = 33$ MeV,
and the nuclear matter compression modulus $K_{nm} = 230$ MeV. 

In Ref.~\cite{Nik.15} we analysed the stability of model parameters of the functional 
DD-PC1 in nuclear matter, and determined the weakly and strongly constrained combinations 
of parameters. In particular, employing a set of pseudo-observables 
in infinite and semi-infinite nuclear matter, the behavior of the cost function $\chi^2$ around the best-fit   
point was analysed. Uncertainties of model parameters and correlation coefficients 
between parameters were computed, as well as the eigenvectors and eigenvalues of the 
Hessian matrix (second derivatives of $\chi^2$) at the best-fit point. It was shown that in addition to combinations of model 
parameters that are firmly constrained by nuclear matter pseudo-data (stiff directions in parameter space), 
soft directions in parameter space correspond to small eigenvalues of the Hessian, 
and represent combinations of model parameters that are only weakly or not at all constrained in nuclear matter. 
It was pointed out that the adopted ansatz for the density dependence of the coupling parameters of DD-PC1 
should therefore be re-examined. In addition, in Ref.~\cite{Nik.15} we also explored uncertainties of 
observables that were not included in the calculation of the cost function: binding energy of asymmetric nuclear matter, 
surface thickness of semi-infinite nuclear matter, and binding energies and charge radii of finite nuclei.

The behaviour of DD-PC1 is characteristic of a large class of nuclear energy density functionals 
that are very sensitive to only a few combinations of parameters (stiff parameter combinations), 
and respond only weakly to all other (soft) parameter combinations \cite{BS.03,Wat.06,Transtrum.11}. 
Using concepts of information geometry and considering DD-PC1 as an illustrative example, in this work we will 
show how such a functional can be reduced to a simpler effective model with fewer parameters. 

%============================================================
\section{\label{sec-theoretical}DD-PC1 in symmetric nuclear matter}
%============================================================

A complex macroscopic model designed to describe certain physical phenomena and reproduce data, usually 
contains a number of parameters that encode the underlying microscopic degrees of freedom. Although their 
values may be constrained to a certain degree by prior information, that is, by the microscopic theory, 
the parameters must be ultimately calibrated by data, for instance, in a nonlinear least-squares fit. 
In a least-squares fit one assumes that the mathematical form of the model (in our case the density 
functional) and the distribution of experimental (data) uncertainties are known, while the parameters must be 
inferred from data. 

Let N be the number of data points (observables ${\mathcal{O}}_n$), and we assume that the model depends 
on F dimensionless parameters $\mathbf{p}=\{p_1,\dots,p_F\}$. The model is interpreted as a manifold of dimension F embedded 
in the Euclidean data space $\mathbb{R}^N$, and the parameters are coordinates for the manifold \cite{Transtrum.11}. If 
each data point ${\mathcal{O}}_n$ is generated by the parameterized model plus random Gaussian noise, 
maximizing the log-likelihood corresponds to minimizing the cost function $\chi^2(\mathbf{p})$: 
\begin{equation}
  \chi^2(\mathbf{p}) 
  = \sum_{n=1}^N{ r_n^2(\mathbf{p}) },
\end{equation}
where $r_n(\mathbf{p})$ denotes the residual
\begin{equation}
r_n(\mathbf{p}) = \frac{\mathcal{O}_n^{(mod)}(\mathbf{p})-\mathcal{O}_n}
                         {\Delta \mathcal{O}_n} ,
\end{equation}
and $\mathcal{O}_n^{(mod)}$ are model predictions that depend on the set of parameters $\mathbf{p}=\{p_1,\dots,p_F\}$.
Every observable is weighted by the
  inverse of $\Delta \mathcal{O}_n$. When
  calibrating a model one often uses an ``adopted error'' that  
  is supposed to include all sources of uncertainty and is adjusted in
  such a way that $\chi^2(\mathbf{p}_0)\approx N-F$ for the
optimal set $\mathbf{p}_0$. The ``best model'' corresponds to the minimum of 
$\chi^2$ on the model manifold, that is, the manifold of model predictions embedded in the data space:
\begin{equation}
\left.\frac{\partial \chi^2(\mathbf{p}) }{\partial p_\mu}\right|_{\mathbf{p}=\mathbf{p}_0}
=0,\quad \forall \; \mu=1,\dots ,F.
\end{equation}
Points in the parameter space are denoted by Greek letters, while Latin letters refer to points in the data space. 
The Hessian matrix of second derivatives ${\partial^2 \chi^2 }/{(\partial p_{\mu} \partial p_{\nu})}$
is positive-definite at  $\mathbf{p}_0$. In the quadratic approximation of the cost function $\chi^2$ around the 
best-fit point: 
\begin{equation}
  \Delta \chi^2(\mathbf{p}) = 
  \chi^2(\mathbf{p}) - \chi^2(\mathbf{p}_0)  =
  \frac{1}{2} \Delta \mathbf{p}^T \hat{\mathcal{M}} \Delta \mathbf{p}\;,
\end{equation} 
where
\begin{equation}
  \mathcal{M}_{\mu \nu}  = 
   \left. \frac{\partial^2 \chi^2 }{\partial p_\mu \partial p_\nu} \right|_{\mathbf{p}=\mathbf{p}_0},
      \label{M} 
\end{equation}
and  $\Delta \mathbf{p}    = \mathbf{p} - \mathbf{p}_0$. 
The  curvature matrix $\hat{\mathcal{M}}$ is symmetric and can be
diagonalized by an orthogonal transformation: 
$\hat{\mathcal{M}}=\hat{\mathcal{A}}\hat{\mathcal{D}}\hat{\mathcal{A}}^T$,
where $\hat{\mathcal{A}}$ denotes the orthogonal matrix with columns
corresponding to normalized eigenvectors of $\hat{\mathcal{M}}$, and 
the diagonal matrix $\hat{\mathcal{D}}$ contains the 
eigenvalues of $\hat{\mathcal{M}}$.  The deviation of 
$\chi^2$ from its minimum value can be expressed as 
\begin{equation}
  \Delta \chi^2(\mathbf{p}) 
  = \frac{1}{2}
  \Delta \mathbf{p}^T \left( \mathcal{A}\mathcal{D}\mathcal{A}^T  \right)
   \Delta \mathbf{p} 
  = \frac{1}{2}
  \mathbf{\xi}^T \mathcal{D} \mathbf{\xi} = \frac{1}{2} \sum_{\alpha =1}^F{\lambda_\alpha
    \xi_\alpha^2} .
\label{delta_chi}
\end{equation}
%------------------
The transformed vectors $\mathbf{\xi}=\hat{\mathcal{A}}^T\mathbf{p}$ define 
the principal axes on the $F$-dimensional model manifold. The concept of 
sloppiness can be quantified by considering the eigenvalues of the 
Hessian (curvature) matrix at the best-fit point. Generally one observes that for 
sloppy models these egeinvalues, characterising sensitivity to variations along orthogonal 
directions in parameter space, are approximately evenly spaced in log-space, extending 
over many orders of magnitude \cite{BS.03,Wat.06,Gut.07,Transtrum.10,Machta.13,Transtrum.15}. 
Each parameter combination is less important than the previous one by a fixed factor. The behaviour of the 
model crucially depends on only a few {\em stiff} directions in parameter space 
characterised by large eigenvalues $\lambda_\alpha$, that is, the cost function $\chi^2$
increases rapidly along these directions and the corresponding linear
combinations of parameters are tightly constrained by the data
that determine $\chi^2$. The remaining {\em soft} directions correspond to small 
eigenvalues $\lambda_\alpha$. This means that there is little deterioration in  
$\chi^2$ as the model moves along a direction defined by the 
eigenvector of $\hat{\mathcal{M}}$ that corresponds to a small eigenvalue. 
Soft linear combinations of bare model parameters are poorly
constrained by the data used in the least-squares fit. 

The sensitivity of a model to variations along an eigenvector of $\hat{\mathcal{M}}$ 
in parameter space is determined by the square root of the corresponding eigenvalue. 
Sloppy models exhibit an exponentially large range of sensitivities to changes in 
parameter values. For instance, if the ratio of the largest to the smallest eigenvalue is $10^6$, the 
combination of parameters in the softest directions has to be changed $10^3$ times 
more than in the stiffest direction to induce the same change in the model behavior. 
Obviously there is no sharp boundary between stiff and soft directions. 
The uncertainties, that is, the variances of model parameters are given by the diagonal 
elements of the inverse matrix $\hat{\mathcal{M}}^{-1}$ (the covariance matrix). This 
means that to constrain the softest combination of parameters to the same level 
of accuracy as the stiffest one, one would have to include in the fit additional 
orders of magnitude more data for that particular direction in parameter space. 
In most cases, of course, this is not feasible. It might also not be that important because 
predictions are possible without precise parameter knowledge \cite{Transtrum.15}, and 
uncertainties in model predictions are far more important than uncertainties of model parameters.  
Sloppiness does not simply originate from an insufficiency of data that leads to a 
a trivial model overparametrisation \cite{Wat.06}, rather it indicates the presence of 
low effective dimensionality associated with the few stiff parameter combinations. 
A complete but sloppy model of a physical system can, in principle, then be reduced to a simpler 
effective model of lower dimension in parameter space \cite{Transtrum.14,Transtrum.15}. 

The application of these concepts to nuclear energy density functionals will be 
illustrated here with the example of DD-PC1. To simplify the computational procedure, we will 
consider a set of pseudo-data in symmetric nuclear matter. This allows calculating analytically  
the derivatives of observables with respect to model parameters, and it also illustrates a 
standard semi-empirical procedure of nuclear energy density 
functional construction in which a specific functional 
of the nucleon one-body density is adjusted to reproduce a microscopic nuclear matter equation of state, 
with a further fine-tuning of (additional) parameters to data on finite nuclei.   

Infinite symmetric nuclear matter is the simplest many-body system for which an energy density 
functional makes definite model predictions depending on the choice of parameters. 
The seven parameters of the isoscalar part of the functional defined in Eq.~(\ref{parameters})
are adjusted to a set of pseudo-data listed in Table~\ref{Tab:pseudoobservables}.
In this illustrative example the minimal data set contains seven points of the microscopic 
nuclear matter equation of state of Akmal, Pandharipande and 
Ravenhall~\cite{APR.98}, based on the Argonne $V_{18}NN$ potential and the UIX
three-nucleon interaction. The EoS points span an interval of nucleon density that 
extends up to two times the saturation density. 
Binding energy as a function of density, however, alone is not sufficient to 
determine both the scalar and vector channels of a relativistic EDF.  For this  
we must also include at least the value of the Dirac mass at or close to the saturation point 
(see the Appendix for the definition of Dirac mass).
The particular value $M_D(\rho =0.152\;fm^{-3}) =0.58 M$ ($M$ denotes the bare
nucleon mass) is the one adopted for the functional DD-PC1, and is 
also consistent with most modern relativistic EDFs.  As we are considering pseudo-data, 
to calculate $\chi^2$ and the Hessian matrix a relatively large arbitrary uncertainty of $10\%$ is 
assigned to each point of the EoS while, since the value of the Dirac mass is already the one of DD-PC1, 
for this quantity the adopted uncertainty is $2\%$. The results and our conclusions will, of course, 
not depend on the specific choice of pseudo-data uncertainties.

\begin{figure}[htb]
\centering
\includegraphics[scale=0.5]{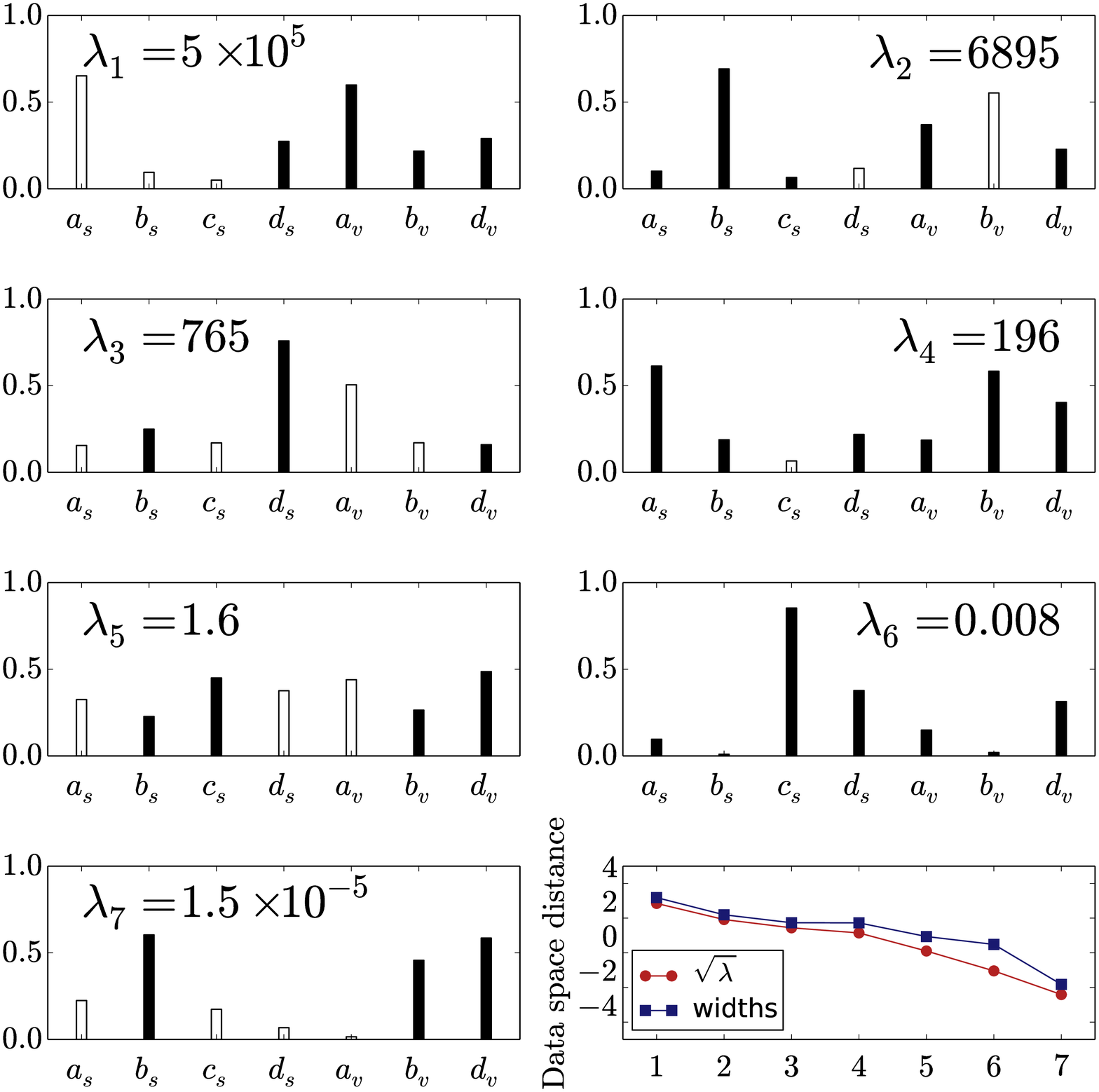} 
\begin {center}
\caption{\label{fig:mat-A} (Color online) Eigenvectors and eigenvalues of the
$7\times 7$ Hessian matrix of second derivatives $\mathcal{M}$ of $\chi^2(\mathbf{p})$ 
at the best-fit point 
in symmetric nuclear matter for the functional defined by the couplings of Eq.~(\ref{parameters}). 
The empty and filled bars indicate that the corresponding amplitudes contribute with 
opposite signs. The last panel displays the logarithmic plot of the widths of the model manifold for each 
of the seven eigendirections, in comparison to the 
square-roots of the corresponding eigenvalues of the Hessian.}
\end{center}
\end{figure}

 %----------------------------------------------------------------------------------------------------------
\begin{table}[t!]
\begin{center}
\caption{\label{Tab:pseudoobservables} Pseudo-data for infinite symmetric nuclear matter
used to compute the cost function $\chi^2$ for the energy density functional defined 
by Eq.~(\ref{parameters}). The seven points correspond to the microscopic EoS of Akmal, 
Pandharipande and Ravenhall~\cite{APR.98}. In the least-squares fit 
the adopted error for the EoS points is 10\%, 
while it is 2\% for the Dirac mass $M_D$.}
\bigskip
\begin{tabular}{c|c} 
\hline
\hline
{\sc pseudo-observable} &  \\ \hline \hline
$\epsilon(0.04\;fm^{-3})$        & -6.48 MeV \\
$\epsilon(0.08\;fm^{-3})$        & -12.43 MeV \\
$\epsilon(0.12\;fm^{-3})$        & -15.43 MeV \\
$\epsilon(0.16\;fm^{-3})$       & -16.03 MeV \\
$\epsilon(0.20\;fm^{-3})$       & -14.99MeV \\
$\epsilon(0.24\;fm^{-3})$       & -12.88 MeV\\
$\epsilon(0.32\;fm^{-3})$       & -6.49 MeV \\ \hline
$M_D(0.152\;fm^{-3})$           &  0.58M \\ \hline \hline
\end{tabular} 
\end{center}
\end{table}
%----------------------------------------------------------------------------------------------------------
The set of all possible values of model
parameters defines the F-dimensional manifold embedded in the N-dimensional data space.
The empirical pseudo-data to which the parameters of the functional are fitted represent a single 
point in the data space, and the best-fit corresponds to the point on the manifold that is 
nearest to the empirical data point.
The Hessian matrix $\mathcal{M}$ of second derivatives of $\chi^2(\mathbf{p})$ 
at the best-fit point $\mathbf{p}_0$ is diagonalized by means of an orthogonal transformation. 
The eigenvalues in decreasing order 
and the components of the corresponding eigenvectors are shown in 
the first seven panels of Fig.~\ref{fig:mat-A}.
Stiff directions in the parameter space are characterised by large eigenvalues. 
Small eigenvalues, on the other hand, refer to soft  
directions in the parameter space, along which $\chi^2$ exhibits little variation and the 
corresponding linear combinations of parameters display large uncertainties and,
therefore, are irrelevant for the behaviour of the functional. Note that the eigenvalues of the Hessian 
span ten orders of magnitude and this spectrum suggests the existence of a lower-dimensional 
model \cite{Transtrum.15}. To remove the irrelevant parameters and construct a simpler functional 
of lower dimension we employ the Manifold Boundary Approximation Method (MBAM) \cite{Transtrum.14}. 

The $N$-dimensional data space is characterised by Euclidian metric and the square of the distance is
the sum of residuals squared $\displaystyle dr^2 = \sum_m{dr_m}^2$. The Jacobian matrix $J_{m \mu}$
that relates changes in the parameters  $\mathbf{p}$ to changes in the residuals is defined by 
\begin{equation}
d r_m = \sum_\mu{\frac{\partial r_m}{\partial p_\mu} dp_\mu} 
=\sum_\mu{J_{m\mu} dp_\mu}\;,
\end{equation} 
and therefore
\begin{equation}
d r^2 = \sum_{m}{dr_m^2} 
%= \sum_m{\sum_{\mu \nu}J_{m\mu}J_{m\nu}dp_\mu dp_\nu}
 = \sum_{\mu \nu}{(J^T J)_{\mu \nu}dp_\mu dp_\nu} 
 = \sum_{\mu \nu}{g_{\mu \nu} dp_\mu dp_\nu}.  
\end{equation}
The Euclidean metric of data space induces a metric on the model manifold: the Fisher information matrix (FIM) $g = J^T J$. 
It can easily be shown that close to the best-fit point
the Hessian matrix $\mathcal{M}$ can be approximated by the metric tensor:
\begin{equation}
\mathcal{M}_{\mu \nu} = 
   \left. \frac{\partial^2 \chi^2 }{\partial p_\mu \partial p_\nu} \right|_{\mathbf{p}=\mathbf{p}_0}
   = \left.\sum_m{\frac{\partial r_m}{\partial p_\mu}\frac{\partial r_m}{\partial p_\nu}}
   \right|_{\mathbf{p}=\mathbf{p}_0} + 
   \left.\sum_m{r_m\frac{\partial^2 r_m}{\partial p_\nu^2}}
   \right|_{\mathbf{p}=\mathbf{p}_0}.
\end{equation}
Since the residuals $r_m$ vanish at the best-fit point, the second term can
be neglected to a good approximation  
\begin{equation}
\label{eq:M-metric-tensor}
\mathcal{M}_{\mu \nu} \approx  \left.\sum_m{\frac{\partial r_m}{\partial p_\mu}\frac{\partial r_m}{\partial p_\nu}}
   \right|_{\mathbf{p}=\mathbf{p}_0}, %= \left.g_{\mu \nu}\right|_{\mathbf{p}=\mathbf{p}_0}.
\end{equation}
which means that $\mathcal{M}_{\mu \nu} \approx g_{\mu \nu}$. Note, however, that 
this relation is approximately valid only in the neighborhood of the best-fit point. 

Sloppy models are characterised by an exponential distribution of eigenvalues of the Hessian matrix and, therefore, an 
exponential sensitivity to parameter combinations. This feature alone, however, does not uniquely determine the sloppiness 
of the model \cite{Transtrum.15}. As a parametrisation independent measure of sloppiness, it has been noted that model 
manifolds of nonlinear sloppy models have boundaries \cite{Transtrum.10}, corresponding to points on the 
manifold where the metric becomes singular. Boundaries of model manifolds can be 
analysed using geodesic paths. The arc length of geodesics, along directions specified by the eigenvectors of the Hessian 
matrix at the minimum of $\chi^2$, provide a measure of the manifold width in each of these directions.

At each point of the embedded $F$-dimensional manifold one can define a tangent
space with the basis vectors
\begin{equation}
e_\mu^m = \frac{\partial r_m}{\partial p_\mu},
\end{equation}
where $m\in\{1,\dots,N\}$ denotes the components of the basis vector in the data space. 
A geodesic is defined as a curve whose tangent vectors remain parallel if they are transported along it: 
\begin{equation}
\frac{\partial e^m_\mu}{\partial p_\nu} =\sum_\lambda{ \Gamma_{\mu \nu}^\lambda
   e^m_\lambda} ,
\end{equation}
where the connection coefficients $\Gamma_{\mu \nu}^\lambda$ are defined by the relation
\begin{equation}
\label{eq:connection-coefficients}
\Gamma_{\mu \nu}^\alpha = \sum_{\beta}{(g^{-1})_{\alpha \beta}
\sum_m{ \frac{\partial r_m}{\partial p_\beta}  \frac{\partial^2 r_m}{\partial p_\mu \partial p_\nu}}} \;.
\end{equation}
The parameters corresponding to a geodesic path can be found as the solution of the 
second-order differential equation
\begin{equation}
\label{eq:geodesic-equation}
\ddot{p}_\mu + 
\sum_{\alpha \beta} {\Gamma_{\alpha \beta}^\mu \dot{p}_\alpha \dot{p}_\beta } =0,
\end{equation}
where $\Gamma_{\alpha \beta}^\mu$ are the connection coefficients defined in
Eq.~(\ref{eq:connection-coefficients}), and the dot denotes differentiation with respect
to the affine parametrisation of the geodesic. This differential equation presents an 
initial value problem in the parameter space. Starting from any point on the model 
manifold, one follows the geodesic path in a particular direction until the boundary 
is identified by the metric tensor becoming singular. In our case this 
occurs whenever the residuals become insensitive to changes in the specific 
linear combination of parameters that defines the geodesic, 
and the elements of the corresponding column/row of the metric tensor  
vanish. Alternatively, the metric becomes singular when the set of basis vectors
$e_\mu^m = \partial r_m/\partial p_\mu$ is linearly dependent and 
at least one column/row of the metric tensor can be written as a
linear combination of the other columns/rows.

Starting from the best-fit point $\chi^2(\mathbf{p}_0)$ and the Hessian matrix whose eigenvectors and 
eigenvalues are shown in Fig.~\ref{fig:mat-A}, we have integrated the geodesic equation along the seven 
eigendirections of the Hessian and thus determined the corresponding 
boundaries of the model manifold. 
The connection coefficients, defined in Eq.~(\ref{eq:connection-coefficients}),
contain first and second derivatives of residuals with respect to the
model parameters. These quantities are proportional to the corresponding derivatives
of the pseudo-observables and, in this illustrative example, can be calculated analytically 
(detailed expressions are included in the Appendix).
The geodesic equation is solved as an initial value problem in the parameter space. 
The initial value $\mathbf{p}_{ini}$ corresponds to the best-fit point, while the initial velocities
$\dot{\mathbf{p}}_{ini}$ are determined by the eigenvectors of the Hessian 
at the best-fit point. Because an eigenvector in fact defines two possible
directions for integration (positive and negative), we integrate in both directions 
until a boundary of the manifold is identified~\cite{Transtrum.14}. The sum of the two 
arc lengths equals the width of the manifold for that particular eigendirection. In the last panel of 
Fig.~\ref{fig:mat-A} (panel on the right in the lowest row), we display the logarithms of the 
widths of the model manifold for each of the seven eigendirections, in comparison to the 
logarithms of the square-roots of the corresponding eigenvalues (a measure of the sensitivity 
of the model to the particular combination of model parameters). A remarkable result is that 
the exponential distribution of model manifold widths in the directions of the eigenvectors of the 
Hessian follows very closely the sensitivity of the functional to changes in the values of the 
corresponding parameter combinations. This is a unique characteristic of sloppy 
models \cite{Transtrum.10,Transtrum.11,Transtrum.14,Transtrum.15}, and indicates that there 
is a simpler effective model of lower dimension in parameter space that can equally well reproduce the data set. 

%----------------------------------------------------------------------------------------------------------
%----------------------------------------------------------------------------------------------------------
%----------------------------------------------------------------------------------------------------------
\begin{figure}[htb]
\centering
\includegraphics[scale=0.7]{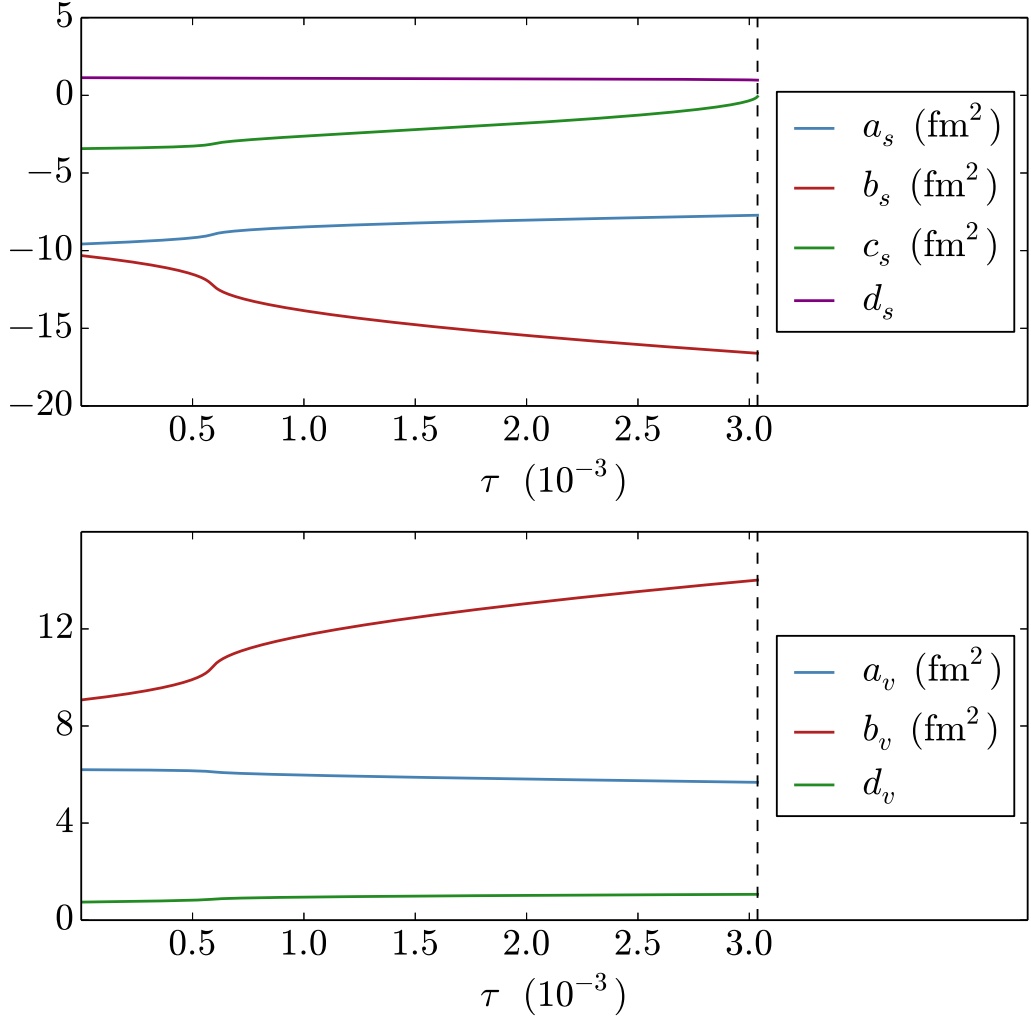} 
\begin {center}
\caption{\label{fig:parameters-step1} (Color online) Evolution of the 
seven parameters of the isoscalar part of the functional defined in 
Eq.~(\ref{parameters}), as functions of the affine parametrisation, along the geodesic path determined 
by the eigenvector of the Hessian matrix that corresponds to the smallest eigenvalue (cf. Fig.~\ref{fig:mat-A}). 
Parameters of the scalar channel are plotted in the upper panel, and the lower panel displays the three 
parameters that determine the vector channel of the functional defined by Eq.~(\ref{parameters}). 
}
\end{center}
\end{figure}
%----------------------------------------------------------------------------------------------------------

%----------------------------------------------------------------------------------------------------------
\begin{figure}[htb]
\centering
\includegraphics[scale=0.5]{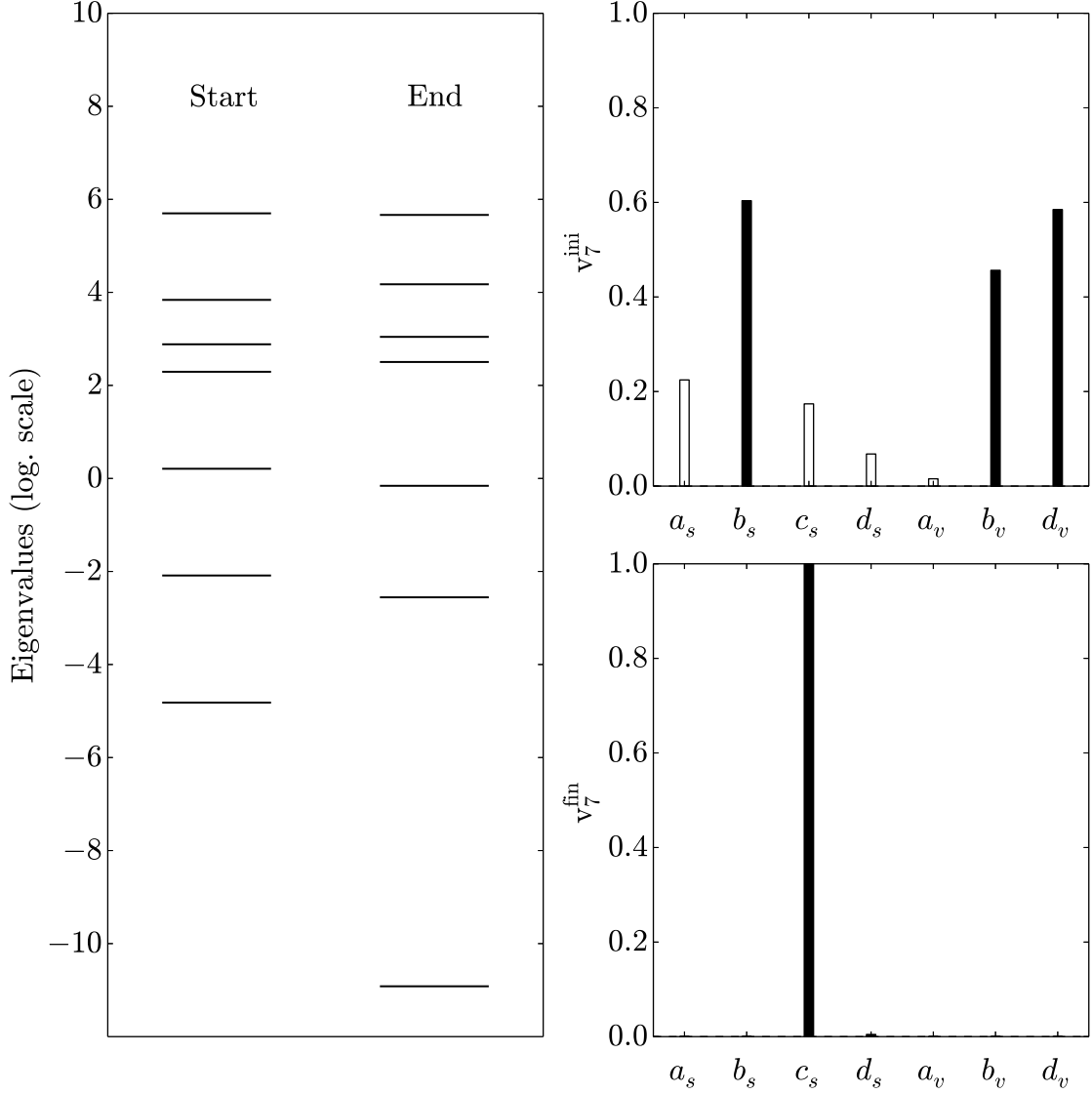} 
\begin {center}
\caption{\label{fig:spectrum-step1} The initial (best-fit point) 
and final (at the boundary of the model manifold) eigenspectrum of the FIM 
for the functional defined by Eq.~(\ref{parameters}), with seven parameters in the isoscalar channel (left panel), and 
the initial and final eigenvectors that correspond to the smallest eigenvalues (panels on the right). 
}
\end{center}
\end{figure}
%----------------------------------------------------------------------------------------------------------

%----------------------------------------------------------------------------------------------------------
\begin{figure}[htb]
\centering
\includegraphics[scale=0.75]{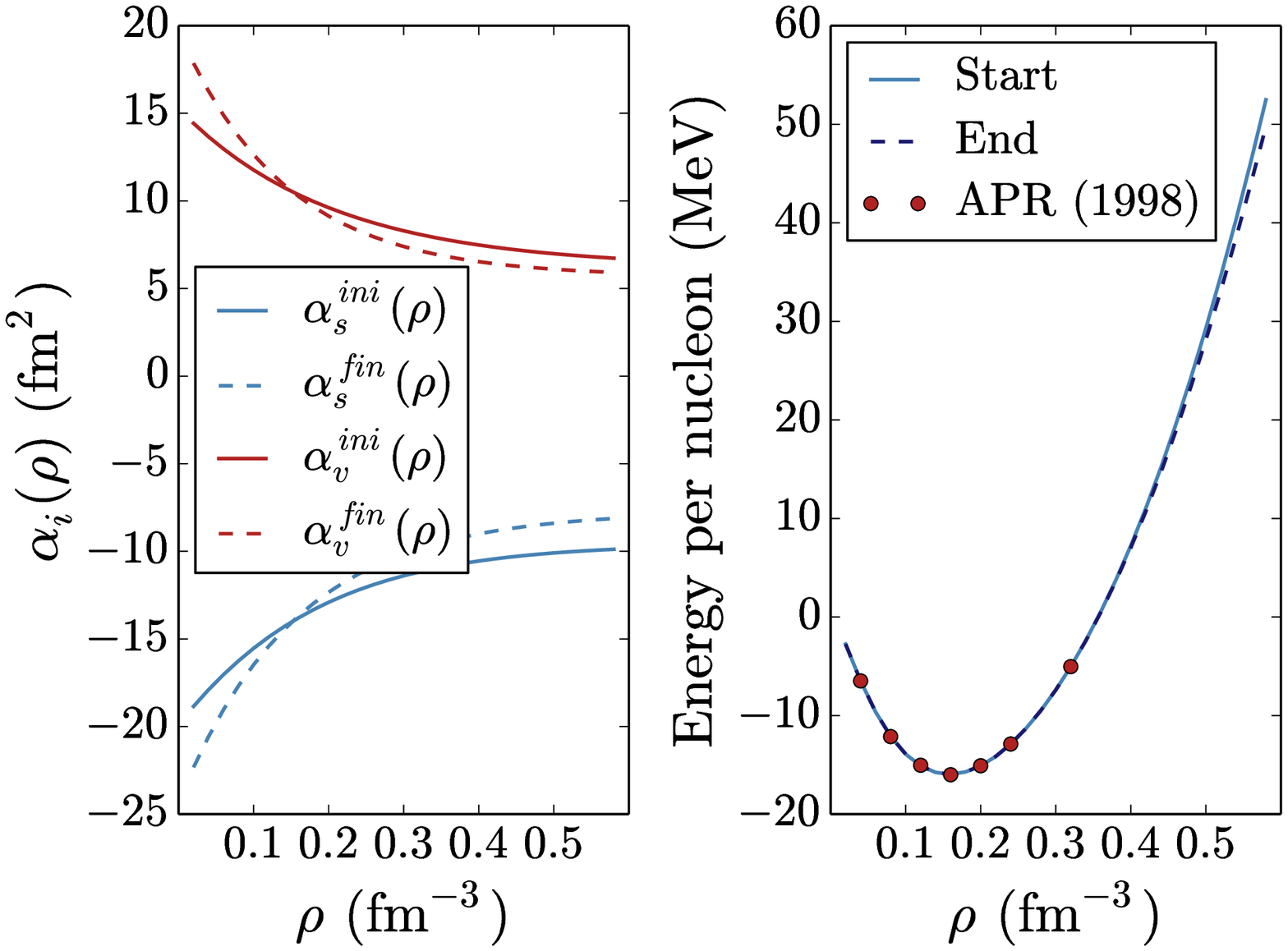} 
\begin {center}
\caption{\label{fig:EoS-step1} (Color online) The initial (best-fit point) 
and final (at the boundary of the model manifold) density-dependent isoscalar coupling 
functions Eq.~(\ref{parameters}) (left), and the corresponding 
initial and final EoS curves (right). Solid curves denote the initial couplings and EoS, while 
dashed curves refer to the couplings and EoS at the boundary. The pseudo-data that 
represent the microscopic EoS are indicated by (red) circles. 
}
\end{center}
\end{figure}
%----------------------------------------------------------------------------------------------------------

%============================================================
\section{\label{MBAM}Reduction of the density functional by manifold boundaries}
%============================================================

Having shown that the functional density dependence of DD-PC1 in fact corresponds to a 
sloppy macroscopic model of the EoS of symmetric nuclear matter, we proceed to 
construct effective models of successively lower dimension until sloppiness is eventually 
eliminated, and all linearly independent parameter combinations are tightly constrained by the 
data set of Table~\ref{Tab:pseudoobservables}. The Manifold Boundary Approximation Method 
(MBAM) \cite{Transtrum.14} essentially consists of four distinct steps. Given a model and a set 
of parameters, in the first step the best-fit parameters are identified, the Hessian matrix 
of the cost function is calculated and diagonalised, 
and the eigendirection with smallest eigenvalue is identified. In the second step the geodesic equation is 
integrated using the parameter values at the best-fit point and the eigendirection with smallest eigenvalue as initial 
conditions, until the boundary of the model manifold is reached. The third step corresponds to the 
evaluation of the model limit associated with this boundary to produce a new model with one less parameters. 
Finally, in the fourth step the new model is optimised by a least-square fit to the data, and used as a starting point 
for the next iteration.

In the first iteration, therefore, we start with the seven parameters of the isoscalar part of the functional defined in 
Eq.~(\ref{parameters}), and the eigenvectors and eigenvalues of the Hessian matrix 
shown in Fig.~\ref{fig:mat-A}. The geodesic equation is numerically integrated with the initial conditions 
described above, and with the additional constraints $\alpha_s(\rho) < 0 $ and $\alpha_v(\rho) > 0$, so that 
the scalar mean-field potential is attractive and the vector mean-field potential repulsive for all values of 
the parameters along the geodesic path (details of the integration of the geodesic equation are given in the
Appendix). The eigenvectors and eigenvalues of the FIM are  
tracked along the geodesic path until the boundary is reached. The behaviour of the parameters along the 
path, as functions of the affine parametrisation of the geodesic, is shown in Fig.~\ref{fig:parameters-step1}.  
Initially the softest eigenvector involves a combination of all the individual bare parameters, and they 
change smoothly along the geodesic. The manifold boundary corresponds to a limit in which one or more 
parameters tend to limiting values (zero or infinity) \cite{Transtrum.14}. In the present case one notes that 
the parameter $c_s$ tends to zero. This is more clearly seen in Fig.~\ref{fig:spectrum-step1} in which we plot 
the initial and final (at the boundary) eigenspectrum of the FIM in the left panel, and the initial and final eigenvectors 
that correspond to the smallest eigenvalues (panels on the right). As the boundary is approached the smallest 
eigenvalue separates from the rest of the spectrum and tends to zero. If the amplitudes of the 
bare parameters in the initial and final eigenvectors are compared, one notes that while initially most parameters 
contribute to the softest eigenvector, at the boundary of the model manifold only the component $c_s$ 
determines the decoupled eigendirection with the eigenvalue of the FIM approaching zero. The initial and final 
coupling functions $\alpha_s(\rho)$ and $\alpha_v(\rho)$, as well as the initial and final curves of the EoS, are 
shown in Fig.~\ref{fig:EoS-step1}. Even though the coupling functions at the boundary of the manifold differ 
considerably from those at the best-fit point, the corresponding EoS curves are virtually indistinguishable and 
reproduce equally well the pseudo-data that represent the microscopic EoS. This is another signature of the 
sloppiness of the model, that is, the model is not sensitive to modifications along 
the softest direction in the parameter space.  

\begin{figure}[htb]
\centering
\includegraphics[scale=0.5]{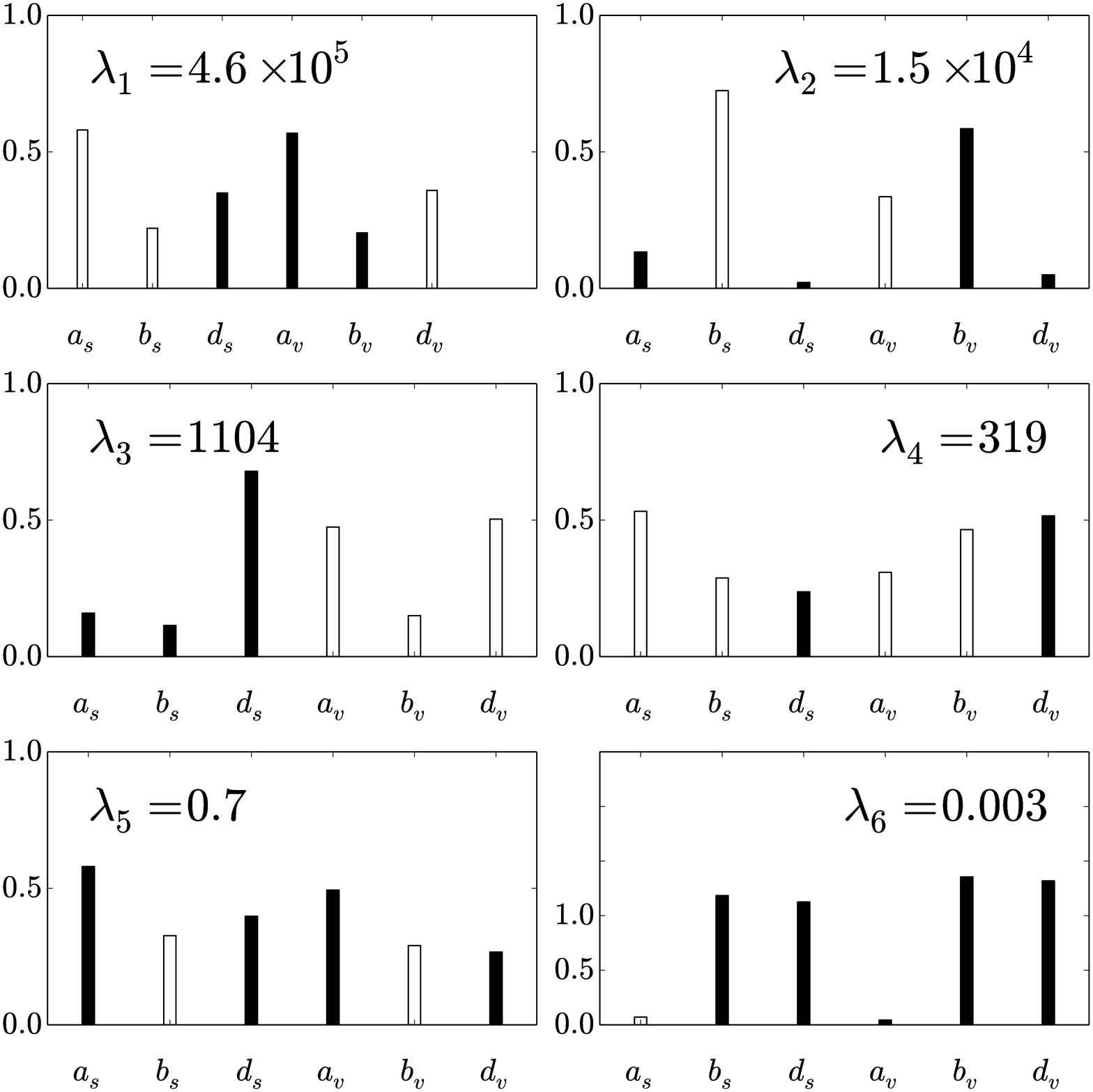} 
\begin {center}
\caption{\label{fig:mat-B} Eigenvectors and eigenvalues of the
$6\times 6$ Hessian matrix of second derivatives $\mathcal{M}$ of $\chi^2(\mathbf{p})$ 
at the best-fit point 
in symmetric nuclear matter for the functional defined by Eq.~(\ref{6_parameters}).
The empty and filled bars indicate that the corresponding amplitudes contribute with 
opposite signs.}
\end{center}
\end{figure}

\begin{figure}[htb]
\centering
\includegraphics[scale=0.7]{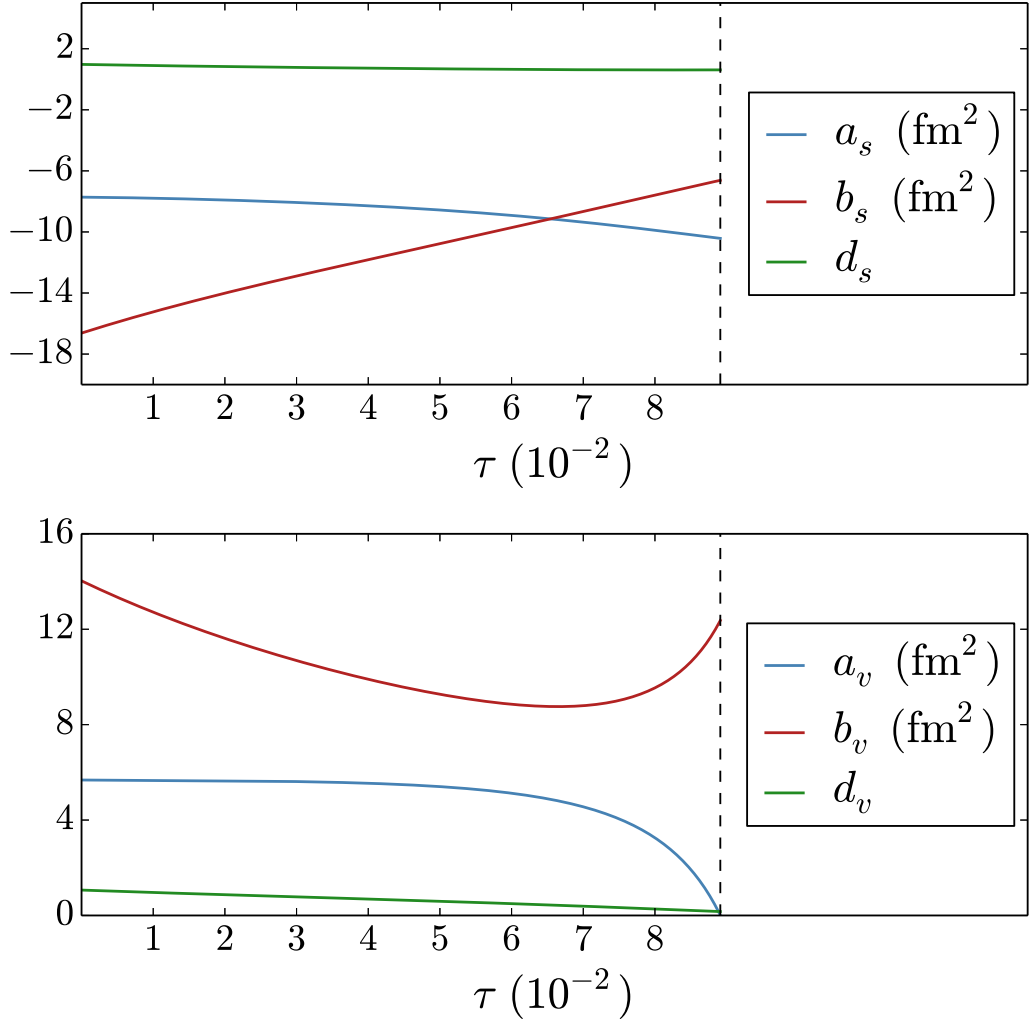} 
\begin {center}
\caption{\label{fig:parameters-step2} (Color online) Same as in the caption to Fig.~\ref{fig:parameters-step1} 
but for the functional defined by the six parameters in Eq.~(\ref{6_parameters}), and the eigendirection for the smallest 
eigenvalue of the Hessian matrix shown in  Fig.~\ref{fig:mat-B}. 
}
\end{center}
\end{figure}

\begin{figure}[htb]
\centering
\includegraphics[scale=0.5]{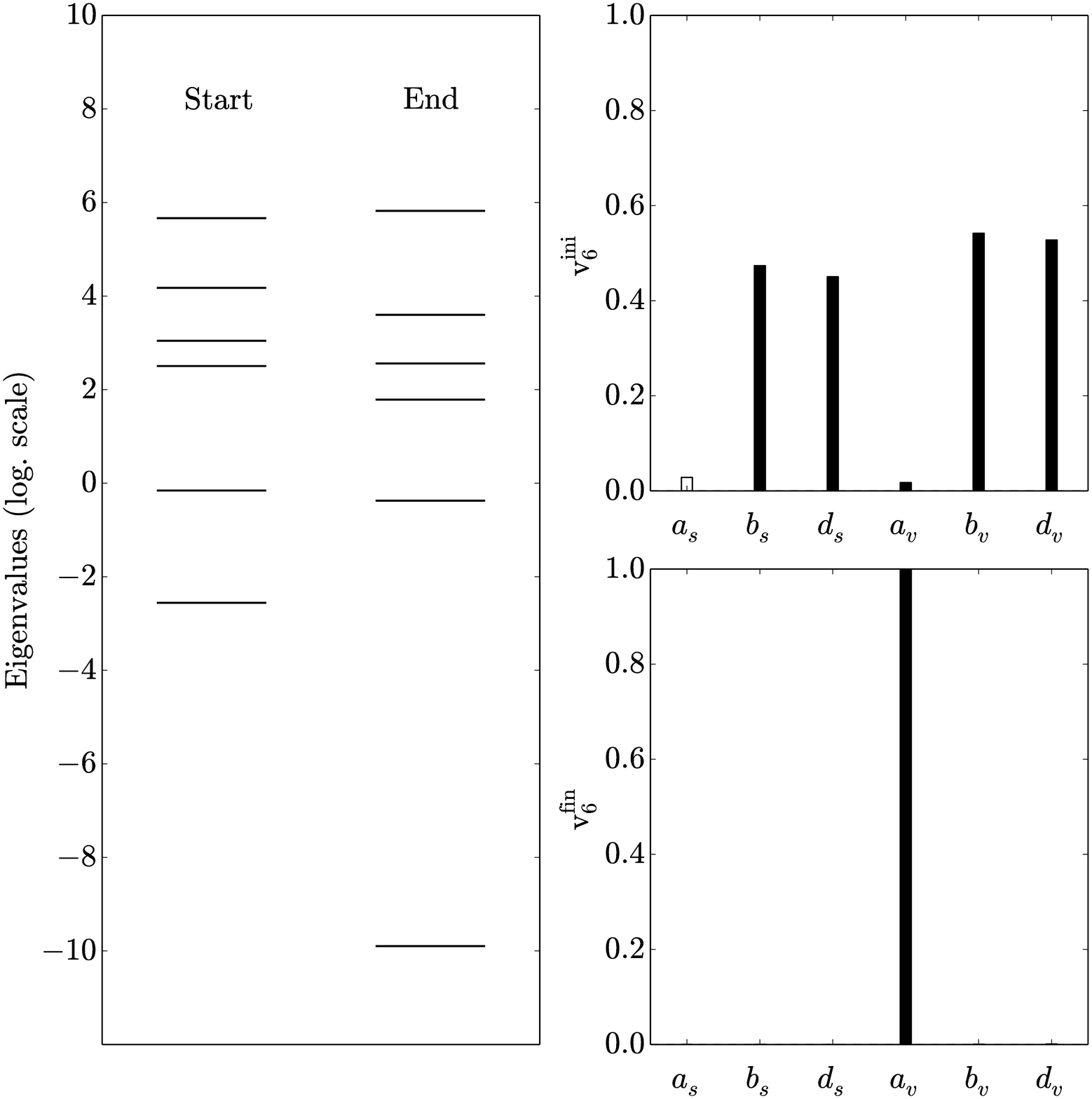} 
\begin {center}
\caption{\label{fig:spectrum-step2} 
Same as in the caption to Fig.~\ref{fig:spectrum-step1} but 
for the functional with six isoscalar parameters defined by Eq.~(\ref{6_parameters}). 
}
\end{center}
\end{figure}

\begin{figure}[htb]
\centering
\includegraphics[scale=0.75]{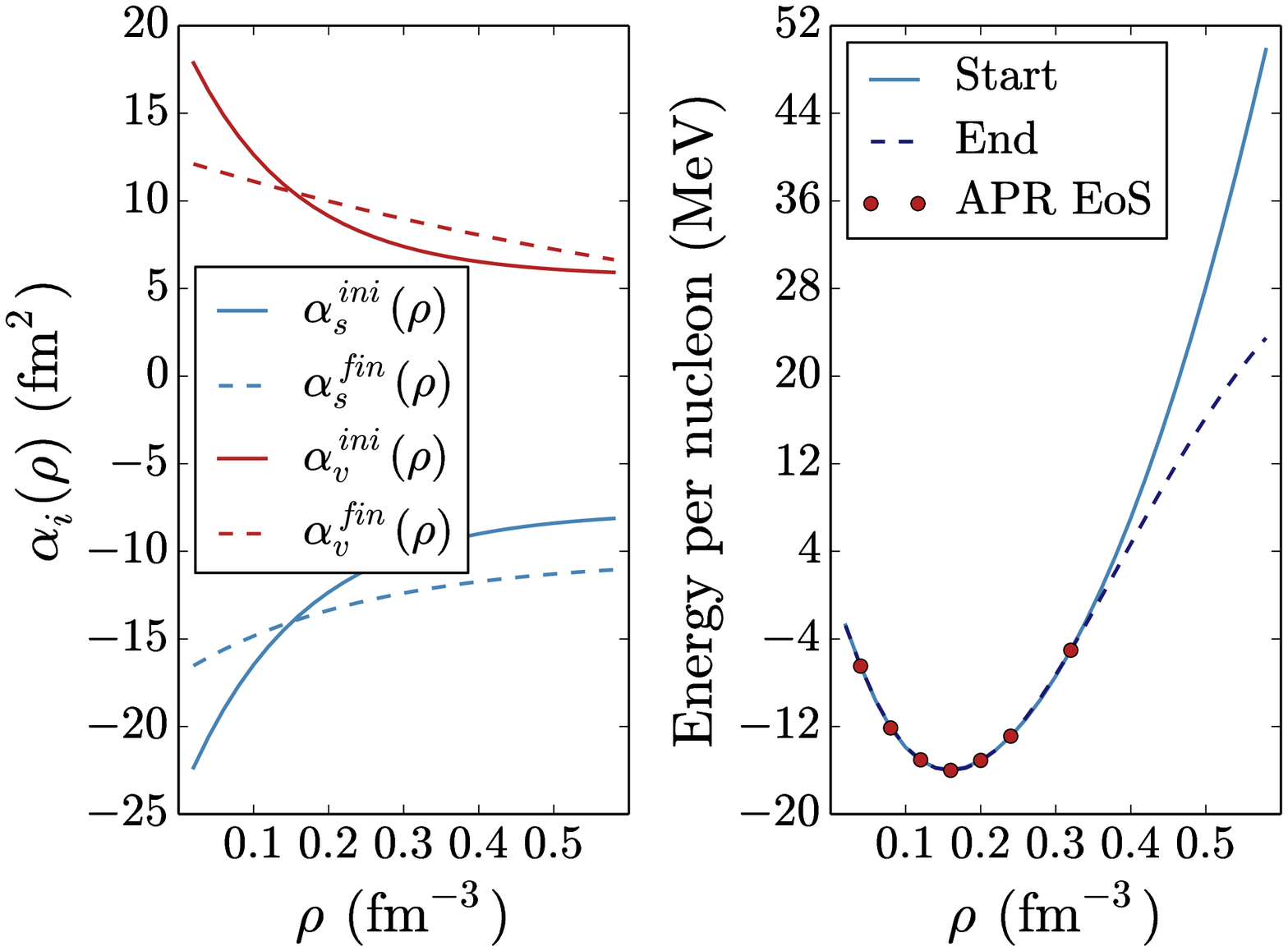} 
\begin {center}
\caption{\label{fig:EoS-step2} (Color online) 
Same as in the caption to Fig.~\ref{fig:EoS-step1} but 
for the functional with six isoscalar parameters defined by Eq.~(\ref{6_parameters}).
}
\end{center}
\end{figure}

From the results shown in Figs. \ref{fig:parameters-step1} - \ref{fig:EoS-step1} one can obviously deduce that
the original seven-parameter model can be reduced to the six-parameter functional form:
\begin{equation}
\alpha_s(\rho) = a_s + b_s e^{-d_s x}, \quad {\rm and} \quad \alpha_v(\rho) = a_v + b_v e^{-d_v x}.
\label{6_parameters}
\end{equation}
In the fourth stage of the first iteration step the new reduced model is readjusted  in a least-squares fit 
to the pseudo-data (Tab.~\ref{Tab:pseudoobservables}), and used as a starting point for the next iteration. 
The corresponding eigenvectors and eigenvalues of the Hessian matrix (FIM at the best-fit point) are shown in 
Fig.~\ref{fig:mat-B}, and here we note that the eigenspectrum spans eight orders of magnitude, compared to 
ten for the model with seven isoscalar parameters. By integrating the geodesic equation with the six best-fit 
initial parameters in the direction of the softest eigenvector, the parameters shown in 
Fig.~\ref{fig:parameters-step2} are obtained. While among the parameters of the scalar coupling only 
$b_s$ changes significantly along the geodesic, it is in the vector channel that parameters take limiting values.
As the boundary of the model manifold is approached, $a_v$ and $b_v$ start diverging but their values are 
limited by the additional constraint on the vector coupling $\alpha_v(\rho) > 0$.
In particular, at the boundary the parameter $a_v$ tends to zero and $d_v$ approaches a small value 
close to zero. The decoupling of the smallest 
eigenvalue at the model manifold boundary is clearly seen in the eigenspectrum of the FIM at the boundary 
(Fig.~\ref{fig:spectrum-step2}), and the eigenvector for the zero eigenvalue is dominated by a single component  
that corresponds to the parameter $a_v$. In the plot of the initial and final 
coupling functions $\alpha_s(\rho)$ and $\alpha_v(\rho)$, and the initial and final curves of the 
EoS (Fig.~\ref{fig:EoS-step2}), one notices that the coupling functions at the boundary display  
a more pronounced difference from the initial ones when compared to the first iteration (cf. Fig.~\ref{fig:EoS-step1}). 
The two corresponding EoS are identical and reproduce equally well the microscopic EoS in the 
interpolating region, as well as the value of the Dirac mass (cf. Tab.~\ref{Tab:pseudoobservables}), 
but start to differ in the region of extrapolation at higher density where no data 
have been specified. 

%----------------------------------------------------------------------------------------------------------
%----------------------------------------------------------------------------------------------------------

%----------------------------------------------------------------------------------------------------------
%----------------------------------------------------------------------------------------------------------
%----------------------------------------------------------------------------------------------------------
%----------------------------------------------------------------------------------------------------------

%----------------------------------------------------------------------------------------------------------
%----------------------------------------------------------------------------------------------------------

The behaviour of the parameters in the vector channel at the boundary (Fig.~\ref{fig:parameters-step2}) suggests 
the following Taylor expansion for the vector coupling function in this limit 
\begin{equation}
\alpha_v(\rho) \approx a_v + b_v(1-d_v x) = a_v +b_v - b_v d_v x = \tilde{a}_v + \tilde{b}_v x.
\end{equation}
The parameters $a_v$ and $b_v$ display opposite asymptotic trends at the boundary, but the 
constraint on the vector coupling prevents $a_v$ from taking negative values. 
In addition, because $d_v$ becomes very small 
at the boundary and $b_v$ asymptotically tends to large values, a single parameter 
$\tilde{b}_v = -b_v d_v$ can be used to parametrise the density dependence of the 
vector coupling function. Therefore, in the next iteration we start with a model determined by 
five parameters: 
\begin{equation}
\alpha_s(\rho) = a_s + b_s e^{-d_s x}, \quad {\rm and} \quad
\alpha_v(\rho) = \tilde{a}_v +\tilde{b}_v x .
\end{equation}

In the third iteration a similar behaviour is observed for the parameters along the geodesic path determined 
by the initial best-fit parameters and the direction of the softest eigenvector, but now for the 
coupling function in the scalar channel of the functional. The parameter $a_s$ tends to zero as 
the boundary of the model manifold is approached (the constraint prevents this parameter from becoming 
positive), while $d_s$ becomes small but finite. 
Performing the Taylor expansion of the scalar function to first order in $d_s$, we take the limit 
$\alpha_s(\rho) = a_s + b_s e^{-d_s x} \approx  a_s + b_s
- b_s d_s x$, and the model can be reduced 
to the four-parameter functional form defined by the coupling functions:
\begin{equation}
\alpha_s(\rho) = \tilde{a}_s + \tilde{b}_s x \quad {\rm and} \quad 
\alpha_v(\rho) = \tilde{a}_v+ \tilde{b}_v x \;.
\label{4_parameters}
\end{equation}

%----------------------------------------------------------------------------------------------------------
\begin{figure}[htb]
\centering
\includegraphics[scale=0.475]{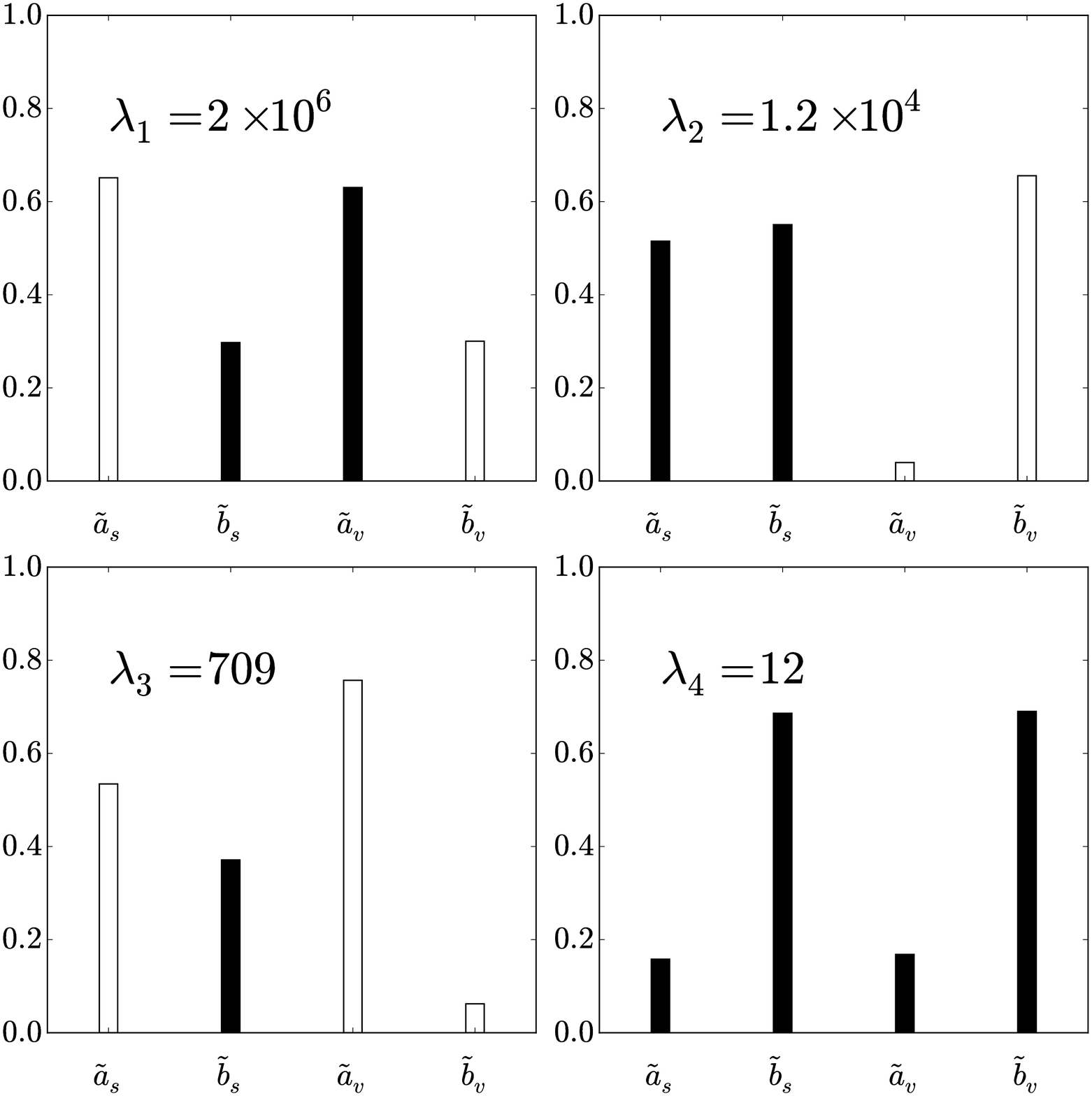} 
\begin {center}
\caption{\label{fig:mat-D} Eigenvectors and eigenvalues of the
$4\times 4$ Hessian matrix of second derivatives $\mathcal{M}$ of $\chi^2(\mathbf{p})$ 
at the best-fit point 
in symmetric nuclear matter for the functional defined by Eq.~(\ref{4_parameters}).
The empty and filled bars indicate that the corresponding amplitudes contribute with 
opposite signs.}
\end{center}
\end{figure}
%----------------------------------------------------------------------------------------------------------

%----------------------------------------------------------------------------------------------------------
\begin{figure}[htb]
\centering
\includegraphics[scale=0.75]{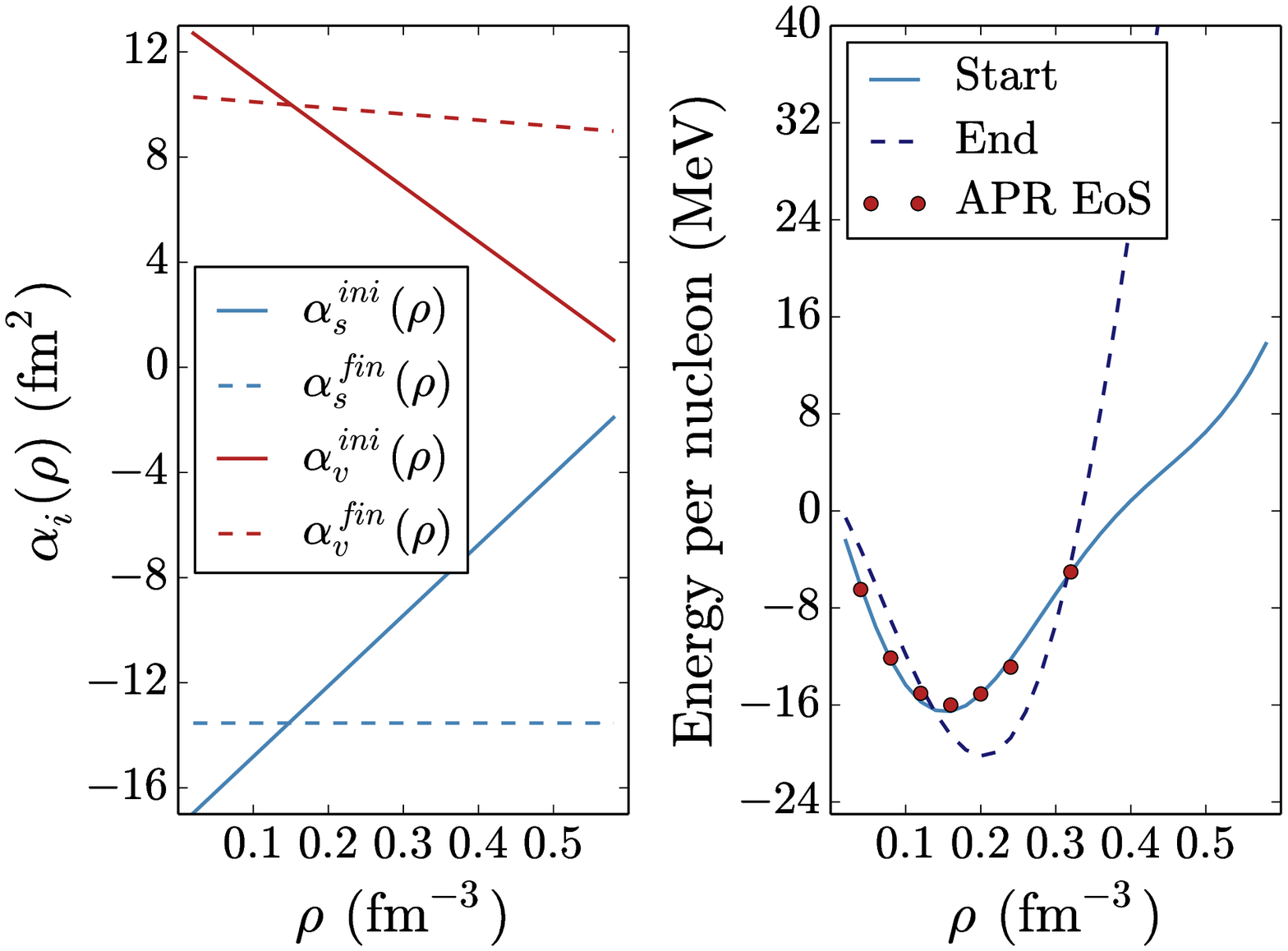} 
\begin {center}
\caption{\label{fig:EoS-step4} (Color online) 
Same as in the caption to Fig.~\ref{fig:EoS-step1} but 
for the functional with four parameters defined by Eq.~(\ref{4_parameters}).
}
\end{center}
\end{figure}
%----------------------------------------------------------------------------------------------------------

When this model is fitted to the microscopic EoS, the resulting eigenvalues and eigenvectors of the 
corresponding Hessian matrix are displayed in Fig.~\ref{fig:mat-D}. We notice that the eigenvalues 
span five orders of magnitude, compared to ten for the original model with seven parameters. 
We could, nevertheless, try to reduce even this model and to that purpose compute the geodesic in the 
direction of the softest eigenvector of the Hessian. 
The result for the corresponding coupling functions $\alpha_s(\rho)$ and $\alpha_v(\rho)$, and the initial and final 
curves of the EoS are shown in Fig.~\ref{fig:EoS-step4}. In this case one notices a very pronounced difference 
between the initial best-fit point and the final point at the boundary, both for the couplings 
as well as for the EoS. The couplings get modified toward constant, density-independent values, while the final EoS displays 
a saturation point at considerably higher density $> 0.2$ fm$^{-3}$, and it is also 
much stiffer. But this means that the model is actually 
very sensitive to parameter variations in the softest direction and, therefore, no longer sloppy. This can also be seen by 
considering the distance in the data space between the best-fit point and the data point, as measured by the 
square root of the cost function. In dimensionless units this distance is: for the model with seven parameters 
0.06, and it does not change when the model is reduced to six parameters. For the model with five parameters 
this distance is 0.11, but it increases to 0.93 when the number of parameter is reduced to four. When we try to 
further reduce the model and parametrise it with only three parameters, the distance between the best-fit point and 
the data point jumps to 5.7, that is, two orders of magnitude more than in the case of six parameters. 
Obviously, as we reduce the complexity of the functional and the model becomes less sloppy, the least-square fit 
deteriorates, initially very little but at some point it becomes unacceptably large. Therefore, one must find a trade-off 
between the sloppiness of the model that will lead to uncertain predictions when extrapolated to regions outside the 
interval of available data, and the agreement with data obtained in a least-square fit.    
%----------------------------------------------------------------------------------------------------------
\begin{figure}[htb]
\centering
\includegraphics[scale=0.75]{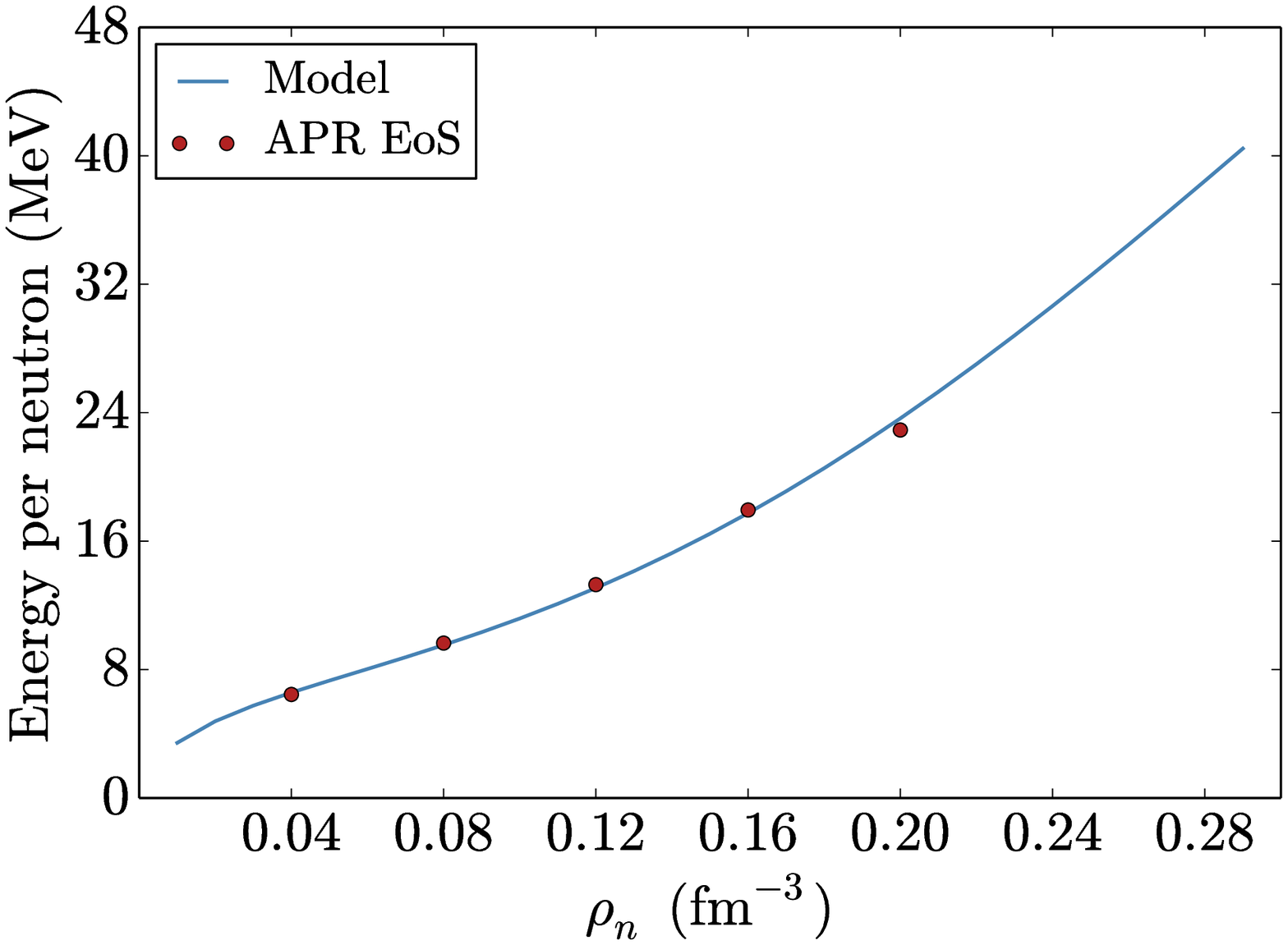} 
\begin {center}
\caption{\label{fig:EoS-NM} (Color online) 
Equation of state of neutron matter, calculated with the isovector channel of the functional DD-PC1 
(solid) in comparison to the pseudo-data that 
represent the microscopic EoS and are indicated by (red) circles. 
}
\end{center}
\end{figure}
%----------------------------------------------------------------------------------------------------------

We note that the functional considered here can, in fact, be further reduced to only two density-independent 
parameters in the isoscalar channel, and still produce an EoS that exhibits saturation at densities 
$\rho \approx 0.2$ fm$^{-3}$. This is, of course, the well known Walecka relativistic mean-field 
model \cite{SW.86,SW.97} which, with just two parameters, in a covariant
treatment of nuclear matter provides a distinction between scalar and
four-vector nucleon self energies, leading to a very natural saturation mechanism. However, as it has been 
known for a long time, without additional density-dependent or higher-order terms that 
phenomenologically take into account many-body in-medium correlations, the Walecka model cannot 
successfully describe ground-state properties of finite nuclei. In the present case, a model with only two 
parameters cannot reproduce the microscopic EoS represented by the pseudo-data listed in 
Tab.~\ref{Tab:pseudoobservables}.

One can also consider the isovector channel of the functional 
(cf. Eq.~(\ref{parameters})). This channel is parametrised by the constants $b_{tv}$ and  $d_{tv}$, 
that can be adjusted to reproduce the corresponding microscopic EoS of neutron matter \cite{APR.98}. 
The binding energy of neutron matter is determined from the following expression:
\begin{equation}
\mathcal{E} = \frac{1}{\pi^2} \int_0^{p_{f,n}}{\frac{p^4 dp}{\left( p^2+M_D^2 \right)^{1/2}} }
+ M (\rho_{s,n} - \rho_{n}) + \frac{1}{2}\alpha_s \rho_{s,n}^2 + \frac{1}{2}\alpha_v \rho_{n}^2 
+ \frac{1}{2}\alpha_{tv} \rho_{n}^2 ,
\end{equation}
where $\rho_{s,n}$ and $\rho_{n}$ denote the scalar and vector (baryon) densities of pure neutron matter, 
respectively, and $p_{f,n}$ is the neutron Fermi momentum.
The result is shown in Fig.~\ref{fig:EoS-NM}, where we plot the EoS of neutron matter calculated with 
the relativistic density functional parametrised with five isoscalar and two isovector parameters. 
Of course, only the two isovector parameters are here specifically adjusted in a least-squares 
fit to the five pseudo-data points that represent the microscopic EoS. A good agreement is obtained and, 
with only two parameters, of course the functional is not sloppy in the isovector channel.

\clearpage

%============================================================
%  Section 5
\section{\label{sec-summary} Summary and outlook}
%--------------------------------------------------------------------------------------------------------
Using concepts from information geometry we have analysed the parameter sensitivity of a nuclear energy density 
functional representative of a large class of semi-empirical functionals currently used in low-energy nuclear 
physics. In a semi-empirical approach a general ansatz is usually adopted for the density dependence of the 
total energy of the nuclear system (infinite nuclear matter and finite nuclei). In a first step most of the parameters are adjusted 
to reproduce a microscopic equation of state (EoS) of symmetric and asymmetric (neutron) matter, while additional 
parameters corresponding to terms in the functional that do not contribute to infinite nuclear matter are directly fitted to 
selected data on finite nuclei. Depending on the number of parameters, semi-empirical functionals have achieved a 
relatively high degree of accuracy in the description of ground-state properties, and have also been successfully employed  
in a number of beyond-mean-field methods, such as the generator coordinate method or the collective Hamiltonian. 
Concerning the predictive power of these functionals and the accuracy of extrapolations to regions where data are 
scarce or not available, however, the situation is far from being satisfactory.

Starting from the density functional form adopted for one of the standard relativistic EDFs: DD-PC1 \cite{NVR.08}, the 
sensitivity of the functional to parameter variations has been analysed in a least-square fit to a microscopic EoS of 
symmetric nuclear matter and neutron matter \cite{APR.98}. In the initial step we have optimised the seven parameters of the 
isoscalar channel of the functional to the EoS of symmetric nuclear matter, and shown that the Hessian matrix of second 
derivatives of the cost function at the best-fit point exhibits an exponential range of sensitivity to parameter modifications. The 
eigenvalues of the Hessian span ten orders of magnitude, ranging from stiff eigendirections in the parameter space that are 
tightly constrained by the set of pseudo-data, to soft eigendirections characterised by very small eigenvalues correponding to 
linear combinations of bare parameters that are poorly determined by data. 

By interpreting the space of model predictions as a manifold embedded in the Euclidean data space, with the parameters of the 
functional as coordinates on the manifold, we have explored the boundaries of the manifold using geodesic paths. 
Starting from the best-fit point, one constructs geodesics along the eigendirections of the Hessian of the cost function, 
and the arc length of the geodesics measures the manifold extension (width) in each of these directions. We have shown that 
the exponential distribution of model manifold widths in the directions of the eigenvectors of the 
Hessian is nearly identical to the distribution of the square roots of the corresponding eigenvalues (sensitivity).  
This is a distinctive signature of sloppy models \cite{Transtrum.10,Transtrum.11,Gut.07,Transtrum.14,Transtrum.15}, that is, 
complex models that can be adjusted to data but are only sensitive to a few stiff parameter combinations, while displaying  
an exponential decrease of sensitivity to variations of soft parameter combinations. In fact, this is a property shared by most 
nuclear energy density functionals that typically use ten or more parameters adjusted to empirical properties of nuclear matter and 
data on finite nuclei. 

A sloppy multi-parameter model can, of course, still be used to make predictions, but its sloppiness really points to an underlying 
model of lower effective dimension associated with the stiff parameters. The reduction of a sloppy model to lower dimension in  
parameter space is, however, a difficult problem and crucially depends on the selected data set. In the second part of this work  
we have employed the Manifold Boundary Approximation Method (MBAM) \cite{Transtrum.14} to simplify and 
deduce the most effective functional form of the density-dependent coupling parameters of our model EDF. We have shown 
that MBAM can indeed be applied to systematically construct simpler nuclear density functionals 
of successively lower parameter space dimension. This is a relatively simple task in nuclear matter, because in this case all 
the derivatives of pseudo-observables with respect to parameters can be calculated analytically, but one must be careful not to 
oversimplify the functional to the point when it is no longer applicable to finite nuclei. We have also found that, as the 
sloppiness of the functional is successively reduced by eliminating soft parameter combinations, the distance in the data space 
between the best-fit point and the data point, measured by the square root of the cost function, increases progressively. 

An interesting problem, therefore, is to find the right balance between the sloppiness of the functional and the level 
of agreement with data obtained by optimising the parameters. As it has been shown in our illustrative example with DD-PC1, 
after a certain number of MBAM iterations the model becomes sensitive to parameter variations in the softest direction. 
Even though the model manifold might still be characterised by boundaries, namely its dimension can in principle be further reduced, 
the resulting best-fit models do not achieve an acceptable agreement with the data point. This is quantified by a pronounced 
increase of the cost function $\chi^2(\mathbf{p}_0)$ for the optimised model with a reduced number of parameters. To construct 
a predictive model, therefore, it becomes more effective to include additional data in the nonlinear least-squares fit that correspond 
to observables ${\mathcal{O}}$ particularly sensitive to soft directions: $\partial {\mathcal{O}} / \partial \xi_{\alpha} >> 1$, 
where $\partial \xi_{\alpha}$ denotes the softest eigendirection in parameter space (cf. Eq.~(\ref{delta_chi})). Covariance analysis 
can then be used to identify the effect of new observables, quantify correlations between different observables predicted by the model, 
and estimate uncertainties of model predictions \cite{Dob14a,Rei10,Fat11}. The inclusion of additional uncorrelated observables in 
the calculation of the cost function can reduce the sloppiness of the model by stiffening formerly soft eigendirections in the 
parameter space and, therefore, improve the performance of the model when extrapolated to regions of the data space not included 
in parameter optimisation. 

This study presents an exploratory analysis that has demonstrated the applicability of methods of information geometry, 
and the MBAM in particular, to the construction and optimisation of nuclear energy density functionals. It suggests that a 
viable strategy is to start with a very general ansatz for the functional form of the density dependence, derived or motivated by 
a microscopic many-body calculation, even if the resulting functional is manifestly sloppy. The complexity and the sloppiness 
of the functional can then be systematically reduced, with the successively smaller set of parameters optimised to both 
empirical properties of nuclear matter, e.g. the EoS, and to selected nuclear data. The latter will, of course, be very 
challenging computationally as the MBAM necessitates the calculation of both first and second derivatives of observables 
with respect to model parameters along geodesic paths on the model manifold. The important result is that, instead of 
{\em a priori} deciding on the form of the functional density dependence to be used in calculations of finite nuclei, 
and then optimising the given set of parameters, by using the MBAM  
it becomes possible that the data that one wishes to describe determine the form of the functional. 
Ideally the final result should be a non-sloppy functional that contains only stiff combinations of parameters and can, 
therefore, be reliably extrapolated to regions where no data are available. 
%----------------------------------------------------------------------------------------------------------------
\begin{acknowledgements}
We would like to thank M. K. Transtrum for helping us with the numerical implementation of the MBAM. 
This work has been supported in part by 
the Croatian Science Foundation -- project ``Structure and Dynamics
of Exotic Femtosystems" (IP-2014-09-9159) and by the QuantiXLie Centre of Excellence.
\end{acknowledgements}
%--------------------------------------------------------------------------------------------------------------------------------------
\appendix
\section{\label{Appendix-formulas} Symmetric nuclear matter}

The expression that relates the baryon (vector) density of symmetric nuclear matter and 
the Fermi momentum reads
\begin{equation}
\label{eq:rho_b}
\rho = \frac{2}{3\pi^2} p_f^3.
\end{equation}
The scalar an vector couplings are assumed to be functions of the baryon density
and furthermore these functions are parametrised by the model parameters
$p_1,\dots,p_n$ in the scalar sector, and $q_1,\dots,q_n$ in the vector sector:
$\alpha_s = \alpha_s(p_1,\dots,p_n;\rho)$ and 
$\alpha_v = \alpha_v(q_1,\dots,q_n;\rho)$.
%--------- 
 \subsection{Derivatives of the Dirac mass}
 The Dirac mass is defined by the following relation
 \begin{equation}
\label{eq:Dirac-mass}
M_D = M + \alpha_s(p_1,\dots,p_n;\rho_B)  \rho_s,
\end{equation}
with the scalar density 
\begin{equation}
\label{eq:rho_s}
\rho_s = \frac{2 M_D}{\pi^2} \int_0^{p_f}{\frac{p^2dp}{\sqrt{{M_D}^2 + p^2}}}.
\end{equation}
The derivative of the Dirac mass with respect to the parameter $p_i$ reads
\begin{equation}
\label{eq:M_D_derivative}
\frac{\partial M_D}{\partial p_i} = \frac{\partial \alpha_s}{\partial p_i}
+ \alpha_s \frac{\partial \rho_s}{\partial M_D}\frac{\partial M_D}{\partial p_i}
\Longrightarrow
\frac{\partial M_D}{\partial p_i} = \frac{\partial \alpha_s}{\partial p_i}
 \frac{1}{1 - \alpha_s \frac{\partial \rho_s}{\partial M_D} }, 
\end{equation}
and the derivative of the scalar density with respect to the Dirac mass
\begin{equation}
\label{eq:rho_s_derivative}
\frac{\partial \rho_s}{\partial M_D} = \frac{\rho_s}{M_D} -
\frac{2M_D^2}{\pi^2}\int_0^{p_f}{\frac{p^2 dp}{\left( p^2+M_D^2 \right)^{3/2}} }.
\end{equation}
The second derivative of the Dirac mass with respect to the parameters $p_i$ and $p_j$ 
is calculated from the expression 
\begin{align}
\frac{\partial^2 M_D}{\partial p_i \partial p_j} &= \frac{\partial}{\partial p_j}
\frac{\partial \alpha_s}{\partial p_i}
 \frac{1}{1 - \alpha_s \frac{\partial \rho_s}{\partial M_D} } \nonumber \\
 &= \frac{\partial^2 \alpha_s}{\partial p_i \partial p_j} \frac{1}{1 - \alpha_s \frac{\partial \rho_s}{\partial M_D} }
  + \frac{\partial \alpha_s}{\partial p_i} \frac{1}{\left( 1- \alpha_s \frac{\partial \rho_s}{\partial M_D} \right)^2}
  \left(   \frac{\partial \alpha_s}{\partial p_j}\frac{\partial \rho_s}{\partial M_D}
  +\alpha_s \frac{\partial^2\rho_s}{ \partial M_D^2}\frac{\partial M_D}{\partial p_j}
  \right), 
\end{align}
and correspondingly the second derivative of the scalar density with respect to the Dirac mass reads
 \begin{equation}
 \frac{\partial^2 \rho_s}{\partial M_D^2 } = \frac{\partial \rho_s}{\partial M_D} \frac{1}{M_D}
 -\frac{\rho_s}{M_D^2}-\frac{4M_D}{\pi^2} \int_0^{p_f}{\frac{p^2 dp}{\left( p^2+M_D^2 \right)^{3/2}} }
 + \frac{6M_D^3}{\pi^2}\int_0^{p_f}{\frac{p^2 dp}{\left( p^2+M_D^2 \right)^{5/2}} } .
 \end{equation}
%---------
\subsection{Derivatives of the binding energy}
The binding energy of symmetric nuclear matter is determined by the relation
\begin{equation}
\mathcal{E} = \frac{2}{\pi^2} \int_0^{p_f}{\frac{p^4 dp}{\left( p^2+M_D^2 \right)^{1/2}} }
+ M (\rho_s - \rho) + \frac{1}{2}\alpha_s \rho_s^2 + \frac{1}{2}\alpha_v \rho^2, 
\end{equation}
where $p_{f}$ denotes  the Fermi momentum. 
The vector coupling $\alpha_v(q_1,\dots,q_n;\rho)$ appears only in the last term and, 
therefore this will be the only term that contributes to the
derivatives with respect to the parameters $q_i$:
\begin{equation}
\frac{\partial \mathcal{E}}{\partial q_i} = \frac{1}{2} \frac{\partial \alpha_v}{\partial q_i}\rho^2,
\quad \frac{\partial^2 \mathcal{E}}{\partial q_i \partial q_j}
 = \frac{1}{2} \frac{\partial^2 \alpha_v}{\partial q_i \partial q_j}\rho^2.
\end{equation}
The scalar coupling, on the other hand, appears in all terms with the Dirac mass
or the scalar density. The derivatives of the binding energy with respect to the
parameters $p_i$ read
\begin{equation}
\frac{\partial \mathcal{E}}{\partial p_i} = 
\left( - \frac{2M_D}{\pi^2}\int_0^{p_f}{\frac{p^4 dp}{\left( p^2+M_D^2 \right)^{3/2}} } 
+ M \frac{\partial \rho_s}{\partial M_D} +\alpha_s \rho_s \frac{\partial \rho_s}{\partial M_D}
\right)\frac{\partial M_D}{\partial p_i}
+ \frac{1}{2}\frac{\partial \alpha_s}{\partial p_i},
\end{equation}
where $\partial \rho_s/\partial M_D$ and $\partial M_D/\partial p_i$ are given in
Eqs. (\ref{eq:rho_s_derivative})  and (\ref{eq:M_D_derivative}) .
Although second derivatives of the binding energy with respect to 
the parameters $p_i$ are more involved, all the necessary expressions
can still be calculated analytically 
\begin{align}
\frac{\partial^2 \mathcal{E}}{\partial p_i \partial p_j} &=
\left( \frac{2}{\pi^2} \int_0^{p_f}{\frac{p^4(2M_D^2-p^2) dp}{\left( p^2+M_D^2 \right)^{3/2}} } 
+ M_D \frac{\partial^2\rho_s}{\partial M_D^2} 
+ \alpha_s \left(\frac{\partial \rho_s}{\partial M_D} \right)^2 \right) 
\frac{\partial M_D}{\partial p_i} \frac{\partial M_D}{\partial p_j} \nonumber \\
&+\left( - \frac{2M_D}{\pi^2}\int_0^{p_f}{\frac{p^4 dp}{\left( p^2+M_D^2 \right)^{3/2}} } 
+ M \frac{\partial \rho_s}{\partial M_D} +\alpha_s \rho_s \frac{\partial \rho_s}{\partial M_D}
\right)\frac{\partial^2 M_D}{\partial p_i\partial p_j} \nonumber \\
&+ \rho_s \frac{\partial \rho_s}{\partial M_D} \frac{\partial \alpha_s}{\partial p_j}
\frac{\partial M_D}{\partial p_i}
+\frac{1}{2} \frac{\partial^2 \alpha_s}{\partial p_i \partial p_j}, 
\end{align}
and we note that the mixed derivatives identically vanish
\begin{equation}
\frac{\partial^2 \mathcal{E}}{\partial p_i \partial q_j} = 0.
\end{equation}
%---------
\section{\label{Appendix-geodesic} Solution of the geodesic equation}
We illustrate the method for integrating the geodesic equation (\ref{eq:geodesic-equation}) by
using the model with seven isoscalar parameters defined in Eq.~(\ref{parameters}), which 
corresponds to the first iteration in the reduction of our initial functional (cf. Figs. \ref{fig:mat-A}-\ref{fig:EoS-step1}). 
The original parameters are initially transformed as follows:
\begin{equation}
\label{eq:reparametrization-scalar}
a_s = a_{s,bf} e^{-p_{a_s}}, \quad b_s = b_{s,bf} e^{-p_{b_s}} ,\quad
c_s = c_{s,bf} e^{-p_{c_s}}, \quad d_s = d_{s,bf} e^{-p_{d_s}}, 
\end{equation}
\begin{equation}
\label{eq:reparametrization-vector}
a_v = a_{v,bf} e^{-p_{a_v}}, \quad b_v = b_{v,bf} e^{-p_{b_v}} ,\quad
d_v = d_{v,bf} e^{-p_{d_v}}, 
\vspace{3mm}
\end{equation}
where $a_{i,bf}$, $b_{i,bf}$, $c_{i,bf}$ and $d_{i,bf}$ denote the parameter values at the best-fit point. 
This transformation ensures that (a) all parameters in the geodesic equation are dimensionless, and 
(b) the exponential form prevents the coupling functions $\alpha_s(\rho) < 0 $ and $\alpha_v(\rho) > 0$ 
to change sign along the geodesic path, so that 
the scalar mean-field potential remains attractive and the vector mean-field repulsive for all 
allowed parameter values.  

The geodesic equation corresponds to a set of coupled ordinary differential equations
\begin{equation}
\ddot{p}_\mu + 
\sum_{\alpha \beta} {\Gamma_{\alpha \beta}^\mu \dot{p}_\alpha \dot{p}_\beta } =0,
\end{equation}
where $\Gamma_{\alpha \beta}^\mu$ are the connection coefficients defined in
Eq.~(\ref{eq:connection-coefficients}), and the dot refers to differentiation with respect
to the affine parametrisation $\tau$ of the geodesic. The number of equations corresponds
to the number of model parameters.

As the initial values the best-fit point parameters are used, and this means that the parameters 
$p_\mu$ in Eqs.~(\ref{eq:reparametrization-scalar})
and (\ref{eq:reparametrization-vector}) are set to zero at the initial point: 
\begin{equation}
p_\mu(0) =  0, \quad \mu \in \{ a_s, b_s, c_s, d_s, a_v, b_v, d_v \} .
\end{equation}
The components of the initial velocity are determined by the amplitudes that correspond to the 
softest eigenvector of the Hessian matrix 
($\mathcal{M} = \mathcal{A} \mathcal{D} \mathcal{A}^T$)
of the cost function at the best-fit point:
\begin{equation}
\label{eq:direction1}
\dot{p}_\mu(0) \sim \mathcal{A}_{\mu}^{\mathrm{softest}}.
\end{equation}
The overall normalisation factor is chosen so that the data space norm of the
velocity vector equals one: 
\begin{equation}
\label{eq:direction2}
\sum_{\mu,\nu}{g_{\mu \nu} \dot{p}_\mu(0)  \dot{p}_\nu(0)} =1,
\end{equation}
where $g_{\mu \nu}$ denotes the metric tensor (FIM). Because the eigenvector is defined up to an 
overall phase, Eqs.~(\ref{eq:direction1}) and (\ref{eq:direction2}) still allow to choose one 
of the two directions for the initial velocity. Here we follow the prescription of Ref.~\cite{Transtrum.14}
(see the Supplemental material), and select the direction in which the parameter space norm of
the velocity vector ($\sum_\mu{\dot{p}_\mu^2}$) increases. We also note that selecting the
opposite direction of integration just leads to the opposing boundary of the manifold.
In this particular case, choosing the opposite direction of integration would lead to
the boundary on which the limit of the model is $\alpha_v(\rho) = \tilde{a}_v + \tilde{b}_v x$, and this 
would simply reverse the first and second iterations.  

The geodesic equation is integrated up to the manifold boundary which is identified by monitoring the
eigenvalues of the metric tensor. Since the data space norm of the velocity remains constant (in our
case one) along the geodesic curve, the length of the traversed path in the data space equals the  
maximal value of the parameter $\tau$. After the solution of the geodesic equation has been obtained, 
the dimensionless parameters $p_\mu$ can be transformed back to the original set 
$a_s$, $b_s$, $c_s$, $d_s$, $a_v$, $b_v$ and $d_v$, and their limiting behaviour at the manifold boundary 
analysed.
%========================================================================================================

%--------------------------------------------------------------------------------------

\end{document}